\documentclass[10pt,journal,twocolumn]{IEEEtran} 
\usepackage{wrapfig,booktabs,fancyhdr,amsmath,amsfonts,tabularx}
\usepackage{cite,bm,bbm,amssymb,amsthm,url,multirow,times,enumitem,comment}
\usepackage{mathtools,siunitx,balance,tikz,adjustbox,array,graphicx}
\usepackage[font=footnotesize]{subcaption}
\usepackage[linesnumbered,ruled]{algorithm2e}
\usepackage[colorlinks=true, linkcolor=black, citecolor=blue, urlcolor=blue]{hyperref}
\newcommand{\vb}{\boldsymbol}

\newcommand{\ba}{\begin{array}}
\newcommand{\ea}{\end{array}}
\newcommand{\sinc}{\text{sinc}}
\newcommand{\define}{\triangleq}

\newcommand{\E}[1]{\mathbb{E}\left[ #1 \right]}

\DeclareMathAlphabet{\mathpzc}{OT1}{pzc}{m}{it}

\DeclareMathOperator*{\argmax}{argmax}

\newcommand{\Fc}[1]{\ensuremath{F_{\text{c}#1}}}
\newcommand{\Fs}{\ensuremath{F_\text{s}}}

\newcommand{\N}{\ensuremath{N}}
\newcommand{\Nr}{\ensuremath{N_\text{r}}}
\newcommand{\Tsym}{\ensuremath{T_\text{sym}}}
\newcommand{\hTsym}{\ensuremath{\hat{T}_\text{sym}}}
\newcommand{\Tf}{\ensuremath{T_\text{f}}}
\newcommand{\Fr}{\ensuremath{F_\text{r}}}
\newcommand{\Fh}{\ensuremath{F_h}}
\newcommand{\Nsf}{\ensuremath{N_\text{sf}}}
\newcommand{\hNsf}{\ensuremath{\hat{N}_\text{sf}}}
\newcommand{\Ndataframe}{\ensuremath{N_\text{sfd}}}
\newcommand{\hNdataframe}{\ensuremath{\hat{N}_\text{sfd}}}
\newcommand{\DeltaFcenter}{\ensuremath{F_\delta}}
\newcommand{\Fg}{\ensuremath{F_\text{g}}}
\newcommand{\T}{\ensuremath{T}}
\newcommand{\F}{\ensuremath{F}}
\newcommand{\Tg}{\ensuremath{T_\text{g}}}
\newcommand{\Ng}{\ensuremath{N_\text{g}}}

\newcommand{\vlos}{\ensuremath{v_\text{los}}}

\newcommand{\Nf}{\ensuremath{N_\text{f}}}
\newcommand{\Ff}{\ensuremath{F_\text{f}}}
\newcommand{\Tfg}{\ensuremath{T_\text{fg}}}
\newcommand{\hTfg}{\ensuremath{\hat{T}_\text{fg}}}
\newcommand{\Td}{\ensuremath{T_\text{d}}}
\newcommand{\Tm}{\ensuremath{T_\text{m}}}
\newcommand{\Tss}{\ensuremath{T_\text{ss}}}
\newcommand{\Mod}{\ensuremath{\:\mathrm{mod}\:}}

\begin{document}

\title{Signal Structure of the Starlink Ku-Band Downlink}

\author{
  \IEEEauthorblockN{Todd E. Humphreys\IEEEauthorrefmark{1},
    Peter A. Iannucci\IEEEauthorrefmark{1},
    Zacharias M. Komodromos\IEEEauthorrefmark{2},
    Andrew M. Graff\IEEEauthorrefmark{2}} \\
  \IEEEauthorblockA{\IEEEauthorrefmark{1}\textit{Department of Aerospace
      Engineering and Engineering Mechanics, The University of Texas at Austin}} \\
  \IEEEauthorblockA{\IEEEauthorrefmark{2}\textit{Department of Electrical and
      Computer Engineering, The
    University of Texas at Austin}}
}

\maketitle

\begin{abstract}
  We develop a technique for blind signal identification of the Starlink
  downlink signal in the 10.7 to 12.7 GHz band and present a detailed picture of
  the signal's structure.  Importantly, the signal characterization offered
  herein includes the exact values of synchronization sequences embedded in the
  signal that can be exploited to produce pseudorange measurements.  Such an
  understanding of the signal is essential to emerging efforts that seek to
  dual-purpose Starlink signals for positioning, navigation, and timing, despite
  their being designed solely for broadband Internet provision. 
\end{abstract}


\newif\ifpreprint
\preprintfalse

\ifpreprint

\pagestyle{plain}
\thispagestyle{fancy}  
\fancyhf{} 
\renewcommand{\headrulewidth}{0pt}
\rfoot{\footnotesize \bf April 2023 preprint of paper submitted for review} \lfoot{\footnotesize \bf
  Copyright \copyright~2023 by Todd E. Humphreys, Peter A. Iannucci, \\
  Zacharias M. Komodromos, and Andrew M. Graff}

\else

\thispagestyle{empty}
\pagestyle{empty}

\fi


\section{Introduction}
In addition to revolutionizing global communications, recently-launched
broadband low-Earth-orbit (LEO) mega-constellations are poised to revolutionize
global positioning, navigation, and timing (PNT).  Compared to traditional
global navigation satellite systems (GNSS), they offer higher power, wider
bandwidth, more rapid multipath decorrelation, and the possibility of stronger
authentication and zero-age-of-ephemeris, all of which will enable greater
accuracy and greater resilience against jamming and spoofing
\cite{Reid2018BroadbandLEO,reid2020leo,kassas2020leo,jardak2022potential,iannucci2022fusedLeo}.

With over 3000 satellites already in orbit, SpaceX's Starlink constellation
enjoys the most mature deployment among LEO broadband providers.  Recent
demonstrations of opportunistic Doppler-based positioning with Starlink signals
\cite{neinavaie2021exploiting,neinavaie2021acquisition,khalife2022first} open up
exciting possibilities.  But whether Starlink signals are more generally
suitable for opportunistic PNT---not only via Doppler positioning---and whether
they could be the basis of a full-fledged GNSS, as proposed in
\cite{iannucci2022fusedLeo}, remains an open question whose answer depends on
details of the broadcast signals, including modulation, timing, and spectral
characteristics.  Yet whereas the orbits, frequencies, polarization, and beam
patterns of Starlink satellites are a matter of public record through the
licensing databases of the U.S. Federal Communications Commission
\cite{starlink2021parameters}, details on the signal waveform itself and the
timing capabilities of the hardware producing it are not publicly available.


We offer two contributions to address this knowledge gap.  First, we develop a
technique for blind signal identification of the Starlink downlink signal in the
10.7 to 12.7 GHz band.  The technique is a significant expansion of existing
blind orthogonal frequency division multiplexing (OFDM) signal identification
methods (see \cite{bouzegzi2010new,gorcin2015ofdm,chaudhari2021design} and the
references therein), which have only been successfully applied to simulated
signals.  Insofar as we are aware, blind identification of operational OFDM
signals, including exact determination of synchronization sequences, has not
been achieved previously.  The technique applies not only to the Starlink
Ku-band downlink but generally to all OFDM signals except as regards some steps
required to estimate synchronization structures that are likely unique to
Starlink.

Second, we present a detailed characterization of the Starlink downlink signal
structure in the 10.7 to 12.7 GHz band.  This applies for the
currently-transmitting Starlink satellites (versions 0.9, 1.0, and 1.5), but
will likely also apply for version 2.0 and possibly later generations, given the
need to preserve backward compatibility for the existing user base.  Our signal
characterization includes the exact values of synchronization sequences embedded
in the signal that can be exploited to produce pseudorange measurements.
Combining multiple pseudorange measurements to achieve multi-laterated PNT, as
is standard in traditional GNSS, enables faster and more accurate opportunistic
position fixes than the Doppler-based positioning explored in
\cite{neinavaie2021exploiting,neinavaie2021acquisition,khalife2022first,psiaki2021navigation}.
and can additionally offer nanosecond-accurate timing, whereas even under the
optimistic scenario envisioned in \cite{psiaki2021navigation}, extracting timing
from Doppler-based processing of LEO signals yields errors on the order of 0.1
to 1 ms.

\section{Signal Capture}
To facilitate replication of our work, and as a prelude to our presentation of
the signal model, we begin with a detailed description of our signal capture
system.

One might reasonably wonder whether a standard consumer Starlink user terminal
(UT) could be modified to capture wideband (hundreds of MHz) raw signal samples
for Starlink signal identification.  Not easily: operating the UT as development
hardware, which would permit capture of raw signal samples, requires defeating
security controls designed specifically to prevent this.  Moreover, the clock
driving the UT's downmixing and sampling operations is of unknown quality and
would therefore taint any timing analysis of received signals.

We opted instead to develop our own system for Starlink signal capture.
Composed of off-the-shelf hardware and custom software, the system enables
signal capture from one Starlink satellite at a time with downmixing and
sampling referenced to a highly-stable GPS-disciplined oscillator.

Whereas the consumer Starlink UT operates as a phased array of many separate
antenna elements, our antenna is a steerable 90-cm offset parabolic dish with a
beamwidth of approximately 3 degrees.  Starlink orbital ephemerides provided
publicly by SpaceX guide our selection and tracking of overhead satellites.
Only one or two Starlink satellites illuminate a coverage cell at any one time
with a data-bearing beam \cite{iannucci2022fusedLeo}.  To guarantee downlink
activity, we solicit data by downloading a high-definition video stream through
a standard Starlink UT co-located with our signal capture system.

\begin{figure*}[t]
  \centering
  \includegraphics[width=\textwidth]{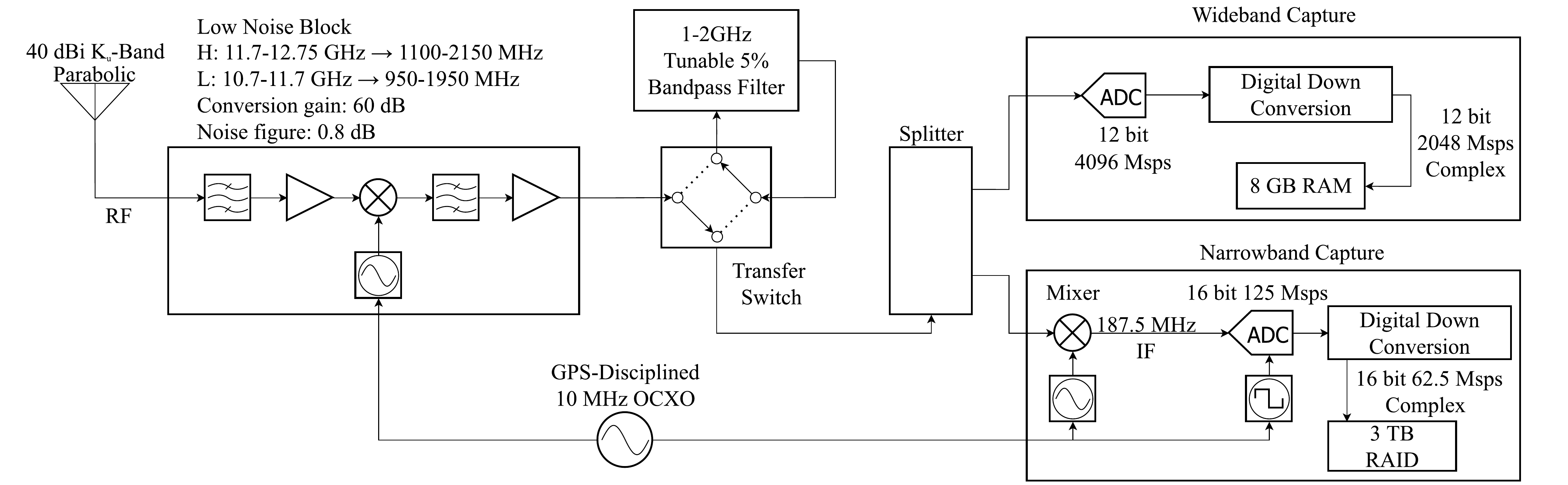}
  \caption{Block diagram of the Starlink signal capture process.}
  \label{fig:sig-cap}
\end{figure*}

Fig. \ref{fig:sig-cap} outlines our signal capture hardware and signal pathways.
A parabolic dish focuses signals onto a feedhorn connected to a low-noise block
(LNB) with a conversion gain of 60 dB and a noise figure of 0.8 dB.  The LNB is
dual-band, downconverting either 10.7--11.7 GHz (the lower band) to 950--1950
MHz, or 11.7--12.75 GHz (the upper band) to 1100--2150 MHz.  The antenna's
nominal gain is 40 dBi at 12.5 GHz, but there are losses of at least 4-5 dB due
to lack of a circular-to-linear polarizer and to feedhorn misalignment.

The signal capture system allows selection between narrowband ($\sim$ 60 MHz)
and wideband ($\sim$ 1 GHz) signal capture modes.  For the narrowband mode, the
output of the LNB is fed to a transfer switch that diverts the signal through a
tunable bandpass filter for image rejection.  Downstream hardware then performs
downmixing (consistent with the selected band), additional bandpass filtering,
and 16-bit complex sampling at 62.5 Msps. The downmixing operation in the LNB
and the downmixing and sampling operations in the downstream hardware are
phase-locked to a common GPS-disciplined oven-controlled crystal oscillator
(OCXO) to minimize the effects of receiver clock variations on the received
signals.  A 3-TB data storage array permits archival of several hours of
continuous data.

Anti-alias filtering prior to sampling reduces the usable bandwidth of the
narrowband mode to approximately 60 MHz.  Although this is much narrower than a
single Starlink channel, multiple overlapping captures can be combined for a
comprehensive analysis of all embedded narrowband structures, as will be shown.
However, the narrowband mode cannot support a synoptic signal analysis.  A
second capture mode---the wideband mode---addresses this deficiency. Based on
direct digital downconversion of 12-bit samples at 4096 Msps (real), the
wideband mode is capable of alias-free capture of the LNB's entire lower band
and most of its upper band.  The wideband mode's limitations are storage,
timing, and noise figure: our current hardware permits only 1-second segments of
contiguous data to be captured before exhausting the onboard memory, the
sampling is not driven by the same clock used for LNB downmixing (due to
hardware limitations), and the noise figure results in captured signals with a
signal-to-noise ratio (SNR) that is significantly worse than for the narrowband
mode.

For the analysis described subsequently, signal identification was based on
narrowband-mode-captured data except for estimation of the primary
synchronization sequence.

\section{Signal Model}
\label{sec:signal-model}
Given its widespread use in wireless communications, one might expect OFDM
\cite{cimini1985analysis, zou1995cofdm, armstrong2009ofdm, ancora2011ofdma,
  proakis2008digital} to be the basis of the Ku-band Starlink downlink.
However, OFDM has historically been avoided in satellite communications systems
because its high peak-to-average-power ratio leads to inefficient transmit power
conversion \cite{jiang2008overview}.  Nonetheless, inspection of the Starlink
power spectrum generated from captured data reveals spectrally-flat frequency
blocks with sharp edges, hallmarks consistent with an OFDM hypothesis.
Proceeding under the assumption of an OFDM model, the problem of general signal
identification narrows to one of identifying the values of parameters
fundamental to OFDM signaling.  This section introduces such parameters as it
presents a generic OFDM signal model and a received signal model.

\subsection{Generic OFDM Signal Model}
\label{sec:generic-ofdm-signal-model}
The serial data sequence carrying an OFDM signal's information is composed of
complex-valued symbols drawn from the set
$\{X_{mik} \in \mathbb{C} : m,i,k \in \mathbb{N}, ~k < \N,~i<\Nsf\}$ at a rate
$\Fs$, known as the channel bandwidth. The subscript $k$ is the symbol's index
within a length-$\N$ subsequence known as an OFDM symbol, $i$ is the OFDM
symbol's index within a length-$\Nsf$ sequence of OFDM symbols known as a frame,
and $m$ is the frame index.  Each symbol $X_{mik}$ encodes one or more bits of
information depending on the modulation scheme (e.g., 1 for BPSK, 2 for 4QAM, 4
for 16QAM, etc.), with higher-order modulation demanding higher SNR to maintain
reception at a given acceptably-low bit-error rate (BER)
\cite{proakis2008digital}.  OFDM is a highly spectrally efficient case of
multicarrier signaling in which each $X_{mik}$ modulates one of $\N$ mutually
orthogonal subcarriers with overlapping spectra.  Let $\T = N/\Fs$ be the
interval over which $N$ information symbols arrive, and $F = \Fs/N = 1/T$ be the
subcarrier spacing, chosen as indicated to ensure subcarrier orthogonality over
the interval $T$. Then the baseband time domain signal produced by the $i$th
OFDM symbol of the $m$th frame is expressed over $0 \leq t < T$ as
\begin{align}
  \label{eq:generic-base}
  x^\prime_{mi}(t) & = \frac{1}{\sqrt{N}}\sum_{k=0}^{\N-1} X_{mik} \exp\left(j2\pi \F{} t k\right)
\end{align}
One recognizes this expression as an inverse discrete Fourier transform,
commonly implemented as an IFFT.  Thus, one can think of each $X_{mik}$ as a
complex-valued frequency-domain coefficient.  To prevent inter-symbol
interference (ISI) arising from channel multipath, OFDM prepends a
cyclically-extended guard interval of length $\Tg = \Ng/\Fs$, called the cyclic
prefix, to each OFDM symbol.  With the addition of the cyclic prefix, the OFDM
symbol interval becomes $\Tsym = T + \Tg$, with $T$ being the useful
(non-cyclic) symbol interval.  Due to the time-cyclic nature of the IFFT, the
prepending operation can be modeled by a simple modification of
(\ref{eq:generic-base}) over $0 \leq t < \Tsym$:
\begin{align}
  \label{eq:generic-symbol}
  x_{mi}(t) & = \frac{1}{\sqrt{N}}\sum_{k=0}^{\N-1} X_{mik} \exp\left(j2\pi
              \F{} (t - \Tg) k\right)
\end{align}
The function $x_{mi}(t)$ is called a time-domain OFDM symbol, or simply an OFDM
symbol when there is little risk of confusion with its frequency-domain
representation.

In all wireless OFDM protocols, subsequences of OFDM symbols are packaged into
groups variously called slots, frames, or blocks.  We will use the term frame
to describe the smallest grouping of OFDM symbols that is self-contained in the
sense that it includes one or more symbols with predictable elements to enable
receiver time and frequency synchronization.  Let $\Nsf$ be the number of OFDM
symbols in a frame, $\Tf \geq \Nsf\Tsym$ be the frame period, and
\begin{align*} 
  g_\text{s}(t) =
  \left\{\ba{ll} 1, & 0 \leq t < \Tsym \\
  0,& \text{otherwise} \ea \right.
\end{align*}
be the OFDM symbol support function.  Then the time-domain signal over a single
frame can be written
\begin{align}
  x_m(t) = \sum_{i = 0}^{\Nsf - 1} x_{mi}(t - i\Tsym)g_\text{s}(t - i\Tsym)
\end{align}
Over an infinite sequence of frames, this becomes
\begin{align}
  \label{eq:generic-full}
  x(t) = \sum_{m \in \mathbb{N}} x_{m}(t - m\Tf)
\end{align}

\subsection{Received Signal Model}
\label{sec:rece-sign-model}
As $x(t)$ passes through the LEO-to-Earth channel and later through the receiver
signal conditioning and discretization operations, it is subject to
multipath-induced fading, noise, Doppler, delay, filtering, and digitization.

In our signal capture setup, the receiving antenna is highly directional,
positioned atop a building with a clear view of the sky, and only used to track
satellites with elevation angles above 50 degrees.  Accordingly, the received
signal's delay spread is negligible---similar to the wooded case of
\cite{cid2015wideband}.  In this regime, the coherence bandwidth appears to be
limited primarily by atmospheric dispersion in the Ku-band, which, as reported
in \cite{hobiger2013correction}, amounts to sub-millimeter delay sensitivity to
dry air pressure, water vapor, and surface air temperature for a 200 MHz-wide
signal.  In view of these favorable characteristics, we adopt a simple additive
Gaussian white noise model for the LEO-to-Earth channel.

Doppler effects arising from relative motion between the satellite and ground
receiver are considerable in the Ku band for the LEO-to-Earth channel.  In fact,
they are so significant that, for a channel of appreciable bandwidth, Doppler
cannot be modeled merely as imposing a frequency shift in the received signal,
as in \cite{bouzegzi2010new,gorcin2015ofdm}, or simply neglected, as in
\cite{chaudhari2021design}.  Instead, a more comprehensive Doppler model is
required, consisting of both a frequency shift and compression/dilation of the
baseband signal.

Let $\vlos$ be the magnitude of the line-of-sight velocity between the satellite
and receiver, modeled as constant over an interval $\Tf$, and let
$\beta \define \vlos/c$, where $c$ is the free-space speed of light.  Note that
lack of frequency synchronization between the transmitter and receiver clocks
gives rise to an effect identical to motion-induced Doppler.  In what follows,
we treat $\beta$ as parameterizing the additive effects of motion- and
clock-error-induced Doppler, and we refer to $\beta$ as the carrier frequency
offset (CFO) parameter.

For an OFDM channel bandwidth $\Fs$, the compression/dilation effects of Doppler
are negligible only if $\beta \Fs T_\text{sync} \ll 1$, where
$T_\text{sync}$ is an interval over which OFDM symbol time synchronization is
expected to be maintained to within a small fraction of $1/\Fs$.  Violation of
this condition causes ISI in OFDM receiver processing as the receiver's discrete
Fourier transform operation, implemented as an FFT, becomes misaligned with
time-domain OFDM symbol boundaries.  In the context of standard OFDM signal
reception, $T_\text{sync}$ may be as short as $\Tsym$, whereas for the signal
identification process described in the sequel,  $T_\text{sync} > \Nsf \Tsym$.

Consider a transmitter in LEO at 300 km altitude, a stationary terrestrial
receiver, elevation angles above 50 degrees, and relative
(transmitter-vs-receiver) clock quality consistent with a
temperature-compensated crystal oscillator.  The resulting $\beta$ is limited to
$|\beta| < 2.5 \times 10^{-5}$.  Suppose $T_\text{sync} = 1$ ms.  Then, to
ensure $\beta \Fs T_\text{sync} < 0.1$, $\Fs$ would be limited to 4 MHz, well
below the Starlink channel bandwidth.  Therefore our Doppler model must include
both a frequency shift and compression/dilation of the baseband signal.

With these preliminaries, we may introduce the  baseband analog received
signal model as
\begin{align}
  \label{eq:received_with_dopp}
  y_a(t) & = x((t - \tau_0)(1-\beta)) \\
  &~~~\times \exp\left(j2\pi\left[\Fc{}(1-\beta) - \bar{\Fc{}}\right](t- \tau_0)\right)
  + w(t) \nonumber
\end{align}
where $\Fc{}$ is the center frequency of the OFDM channel,
$\bar{\Fc{}} \approx \Fc{}$ is the center frequency to which the receiver is
tuned, $\tau_0$ is the delay experienced by the signal along the least-time path
from transmitter to receiver, and $w(t)$ is complex-valued zero-mean white
Gaussian noise whose in-phase and quadrature components each have (two-sided)
spectral density $N_0/2$.  Let the symbols $\{X_{mik}\}$ be scaled such that
$x(t)$ has unit average power over nonzero OFDM symbols.  Then, during such
symbols and within the channel bandwidth $\Fs$, $\text{SNR} = 1/N_0\Fs$.

In a late stage of the signal capture pipeline shown in Fig. \ref{fig:sig-cap},
the analog signal $y_a(t)$ is discretized.  Let $\Fr$ be the receiver's sampling
rate and $h(t)$ be the impulse response of a lowpass prefilter with (two-sided)
3-dB bandwidth $\Fh < \Fr$ and rolloff such that power is negligible for
frequencies $|f| > \Fr/2$, permitting alias-free complex sampling.  Then the
baseband digitized received signal model is
\begin{align}
  \label{eq:received signal model}
  y(n) = \int_{-\infty}^{\infty} h(n/\Fr - \tau) y_a(\tau) ~d\tau, \quad n \in \mathbb{Z}
\end{align}
Note that, strictly speaking, (\ref{eq:received_with_dopp}) and
(\ref{eq:received signal model}) apply only to the narrowband capture mode.
Accounting for the distinct mixing and sampling clocks in the wideband mode
would require a more elaborate model.

\section{Signal Identification Preliminaries}
\label{sec:sign-ident}
Here we summarize and augment the terminology and notation previously introduced
to allow a clear statement of the identification problem to be solved.  Then, to
develop intuition about the solution procedure presented in the following
section, we explain how signal cyclostationarity is exploited to estimate key
signal parameters.

\subsection{Terminology and Parameters of Interest} We assume the frequency
spectrum allocated for a multi-band OFDM signal is divided into OFDM
\emph{channels} within which power spectral density is approximately uniform.
Adjacent channels are separated by \emph{guard bands}.  Each channel is composed
of $\N$ orthogonal \emph{subcarriers} whose spectra overlap.  A
\emph{frequency-domain OFDM symbol} is a vector of $N$ complex-valued
coefficients whose $k$th element modulates the $k$th subcarrier.

The IFFT of a frequency-domain OFDM symbol, when prepended by a \emph{guard
  interval} (cyclic prefix), becomes a \emph{time-domain OFDM symbol}.
Subsequences of such symbols are packaged into \emph{frames} in which one or
more OFDM symbols carry predictable elements, called \emph{synchronization
  sequences}, that enable receiver time and frequency synchronization.  As
transmitted, an OFDM signal's carrier phase remains stable within each frame.
Frames are separated from each other by at least the \emph{frame guard
  interval}.  There may be further logical subframe structure (e.g., slots,
header segments), but these are not addressed in this paper's signal
identification process.

Note that three distinct structures share the term ``guard'': the empty spectrum
between channels (\emph{guard band}), the time between frames (\emph{frame
  guard interval}), and the (cyclic) prefix in a time-domain OFDM symbol
(\emph{OFDM symbol guard interval}).

The OFDM parameters of interest for this paper's signal identification problem
are summarized in Table \ref{tab:parameters-description}.
\begin{table}[t]
  \centering
  \caption{Parameters of Interest}
  \begin{tabular}[c]{ll}
    \toprule
    \multicolumn{2}{c}{Independent Parameters}\\
    \midrule
    $\Fs$        & Channel bandwidth; information symbol rate\\
    $\N$         & Number of subcarriers in bandwidth \Fs{}\\
    $\Ng$        & Number of intervals 1/\Fs{} in an OFDM symbol guard interval\\
    $\Tf$        & Frame period\\
    $\Tfg$       & Frame guard interval\\
    $\Nsf$       & Number of non-zero symbols in a frame\\
    $\Ndataframe$& Number of data (non-synchronization) symbols in a frame\\
    $\Fc{i}$     & Center frequency of $i$th channel \\
  \end{tabular}
  \begin{tabular}[c]{ll}
    \toprule
    \multicolumn{2}{c}{Derived Parameters}\\
    \midrule
    $\T  =\:\scriptstyle \N/\Fs$    & Useful (non-cyclic) OFDM symbol interval\\
    $\Tg =\:\scriptstyle \Ng/\Fs$   & Symbol guard interval \\
    $\Tsym =\:\scriptstyle \T + \Tg$  & OFDM symbol duration including guard interval \\
    $\F =\:\scriptstyle \Fs/\N$     & Subcarrier spacing \\
    $\DeltaFcenter =\:\scriptstyle \Fc{i} - \F_{\text{c}(i-1)}$ & Channel spacing \\
    $\Fg =\:\scriptstyle \DeltaFcenter - \Fs$ & Width of guard band between channels \\
    \bottomrule
  \end{tabular}
  \label{tab:parameters-description}  
\end{table}

\subsection{Problem Statement}
\label{sec:problem-statement}
This paper's blind signal identification problem can be stated as follows: Given
one or more frame-length segments of received data modeled by (\ref{eq:received
  signal model}), estimate the value of the independent parameters listed in
Table \ref{tab:parameters-description} with sufficient accuracy to enable
determination of the symbols $\{X_{mik}\}$ that apply within the captured time
interval over the bandwidth $\Fh$.  Also identify and evaluate any
synchronization sequences present within a frame.

Note that this signal identification problem is more demanding than those
treated in the existing blind OFDM signal identification literature, in five
ways.  First, no prior identification procedures were truly blind: they operated
on simulated signals generated by the researchers themselves.  As will be shown,
simulated signals, which assume independent and identically-distributed (iid)
information symbols $\{X_{mik}\}$, bear characteristics markedly different from
operational OFDM signals. Second, prior studies either neglected Doppler effects
or modeled only a bulk frequency shift arising from Doppler.  Third, the goal of
prior work has been limited to distinguishing known OFDM waveforms from each
other \cite{bouzegzi2010new,chaudhari2021design}, or from single-carrier systems
\cite{gorcin2015ofdm}.  As such, they do not estimate the comprehensive set of
independent parameters required to recover the symbols $\{X_{mik}\}$.  For
example, \cite{bouzegzi2010new} estimates the useful symbol interval $T$ and the
symbol guard interval $\Tg$, but not $\Fs$, $N$, and $\Ng$ independently.
Fourth, existing approaches assume the receiver bandwidth $\Fh$ is wider than
$\Fs$, which is not the case for our narrowband capture mode.  Fifth, prior
studies have not been concerned with identifying and characterizing any
synchronization sequences in OFDM frames.  Yet such sequences are key to
standard OFDM signal processing and are especially important for efforts to
dual-purpose OFDM signals for PNT.

\subsection{Exploiting Signal Cyclostationarity}
A fundamental concept exploited in feature-based signal identification is signal
cyclostationarity \cite{dobre2015signal,bouzegzi2010new}.  While all
communications signals exhibit cyclostationarity, it is especially pronounced in
OFDM signals due to the cyclic prefix present in each OFDM symbol.

\begin{figure}[t]
  \centering
  \includegraphics[width=0.48\textwidth]{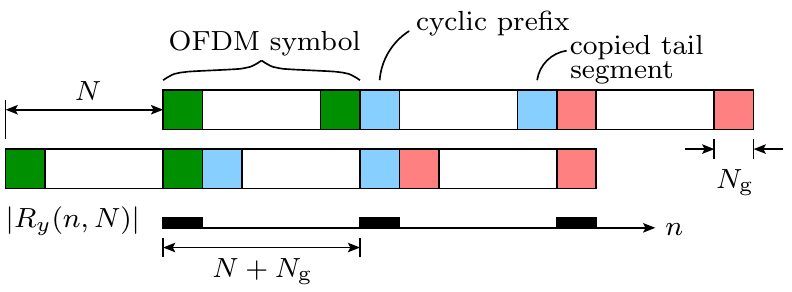}
  \caption{Graphical explanation for why $R_y(n,N)$ is cyclic in $n$ with period
    $N + \Ng$.}
  \label{fig:cyclostationarity}
\end{figure}

To simplify explanations in this subsection, assume that $\beta = 0$, that the
receiver sampling rate $\Fr$ is identical to the OFDM channel bandwidth $\Fs$,
and that the receiver filter bandwidth $\Fh \approx \Fs$.  Then, letting
$\E{\cdot}$ denote the expectation operation, define the autocorrelation
function of the received discrete-time signal $y(n)$ as
\begin{align}
  \label{eq:autocorr-fcn}
  R_y(n,\tau) = \E{y(n+\tau)y^*(n)}
\end{align}
where $y^*(n)$ is the complex conjugate of $y(n)$.  If the coefficients
$\{X_{mik}\}$ are iid and selected randomly from among the possible
constellation values, then $R_y(n,\tau)$ may be nonzero only at
$\tau \in \{0, N, -N\}$ \cite{bouzegzi2010new}. As illustrated in
Fig. \ref{fig:cyclostationarity}, nonzero autocorrelation at
$\tau \in \{N, -N\}$ arises because $y(n)$ is shifted against itself in such a
way that cyclic prefixes are aligned perfectly with the portions of the symbols
of which they are a copy.  Fig. \ref{fig:cyclostationarity} also makes clear
that $R_y(n,N)$ is cyclic in $n$ with period $\N + \Ng$.  Moreover, within a
sequence of nonzero OFDM symbols, $\E{y(n)} = \E{y(n + \N + \Ng)}$.  These
attributes imply that $y(n)$ is wide-sense cyclostationary
\cite{proakis2008digital}.
The autocorrelation function $R_y(n,\tau)$ is the key to determining $N$ and
$\Ng$ without the need for prior time and frequency determination.  Since
$R_y(n,\tau)$ is periodic in $n$ with period $N + \Ng$ for certain values of
$\tau$, it can be expanded in a Fourier series as
\begin{align}
  R_y(n,\tau) = \sum_{\alpha \in \mathcal{A}(\xi)} R_y^\alpha(\tau) \exp\left(j 2 \pi
 n  \alpha \right)
\end{align}
where $\mathcal{A}(\xi) = \{p/\xi : p \in \mathbb{Z}\}$.  The particular set
$\mathcal{A}(N + \Ng)$ contains the so-called cyclic frequencies. The Fourier
coefficient $R_y^\alpha(\tau)$, also called the cyclic autocorrelation function,
equals
\begin{align}
  \label{eq:cyclic_autocorr}
   R_y^\alpha(\tau) = \lim_{M \rightarrow \infty} \frac{1}{M} \sum_{n = 0}^{M-1}
  R_y(n,\tau) \exp\left(-j2\pi n \alpha\right)
\end{align}

Given the nature of $R_y(n,\tau)$, the function $R_y^\alpha(\tau)$ is only
nonzero when $\tau = N$ and when $\alpha$ is one of the cyclic frequencies from
the set $\mathcal{A}(N + \Ng)$.  This fact underlies the following estimators
for $N$ and $\Ng$.  Let $\mathcal{S}$ be the set of possible values of $N$.
Then an estimator for $N$ is obtained as
\begin{align}
  \label{eq:est_N}
  \hat{N} = \argmax_{\tau \in \mathcal{S}} \left| R^0_y(\tau) \right|
\end{align}
Similarly, let $\mathcal{S}_\text{g}$ be the set of all possible values of
$N + \Ng$.  Then, assuming $\hat{N}$ is an accurate estimate of $N$, an
estimator for $\Ng$ is obtained as
\begin{align}
    \label{eq:est_Ng}
  \hat{\Ng} = -\hat{N} + \argmax_{\xi \in \mathcal{S}_\text{g}} \sum_{\alpha \in
  \mathcal{A}(\xi)} \left|R_y^\alpha(\hat{N})\right|
\end{align}
A graphical depiction of the functions being maximized in (\ref{eq:est_N}) and
(\ref{eq:est_Ng}) is provided in the next subsection.  Note that because these
estimators involve an autocorrelation limited approximately to offsets
$|\tau| \leq N$, which amounts to a short time interval of $T = N/\Fs$, they
are robust to nonzero Doppler, provided that $\beta \Fc{} T \ll 1$.

The mathematical structure of these two estimators is similar to the
cyclic-correlation-based method presented in \cite{bouzegzi2010new} except that
they are intended to operate successively rather than jointly, which makes them
more computationally efficient without loss of accuracy.

Observe that both estimators are based on the cyclic autocorrelation function
given in (\ref{eq:cyclic_autocorr}).  In practice, this function is approximated
as
\begin{align}
  \label{eq:cyclic_autocorr_empirical}
  R_y^\alpha(\tau) \approx \frac{1}{M} \sum_{n = 0}^{M-1}
  y(n+\tau)y^*(n) \exp\left(-j2\pi n \alpha\right)
\end{align}
where $M$ is a number much larger than the cyclic period $N + \Ng$, such as the
number of samples in one frame, or even multiple frames if frame-to-frame
correlation is of interest.

\section{Signal Identification Procedure}
\label{sec:sign-ident-proc}
We present here our solution to the signal identification problem posed in
Section \ref{sec:problem-statement}.  To facilitate replication, we present the
solution in the form of a step-by-step procedure.
\subsection{Estimation of  \normalfont  $N$}
We first construct $\mathcal{S}$, the set of possible values of $N$.  Here we
exploit the constraints that designers of OFDM signals must respect when
choosing $N$.  Naturally, they wish to maximize the signal's total data
throughput, which for an OFDM signal with all subcarriers fully modulated is
\begin{align}
  \label{eq:ofdm_throughput}
d_\text{OFDM} = \frac{b_\text{s} \Fs N}{N + \Ng} \quad \text{bits/s}
\end{align}
Here, $b_\text{s}$ is the number of bits per symbol (e.g., 2 for 4QAM
modulation).  Observe that, for given $\Fs$ and $b_\text{s}$, increasing
$d_\text{OFDM}$ implies increasing $N/\Ng$.  But $\Ng$ is lower-bounded by the
physical characteristics of the channel: it must be large enough that
$\Tg = \Ng/\Fs$ exceeds the channel's delay spread.  Thus, designers are
motivated to increase $N$ insofar as possible to maximize throughput.  But they
must respect a practical upper bound on $N$ related to the subcarrier spacing
$F = \Fs/N$: a narrower $F$ puts greater demands on CFO estimation.  Let
$\tilde{\beta}$ be the error in a receiver's estimate of the CFO parameter
$\beta$.  To avoid inter-carrier interference (ICI), which degrades BER,
$\tilde{\beta}$ must satisfy
\begin{align}
  \label{eq:beta_tilde_reqt}
\tilde{\beta}\Fc{} < \epsilon F
\end{align}
where $\Fc{}$ is the OFDM channel's center frequency and $\epsilon$ is limited
to a few percent \cite{ancora2011ofdma}.  Assume that known synchronization
symbols present within a frame allow modulation wipeoff on $\N_\text{sync}$
contiguous samples, exposing the underlying coherent carrier signal from which
$\beta$ can be estimated.  Then a lower bound on the variance of
$\tilde{\beta} \Fc{}$ is given by the Cramér-Rao bound for the frequency
estimation problem with unknown phase and amplitude \cite{rife1974single}:
\begin{align}
  \label{eq:var_beta_Fc}
  \text{var}(\tilde{\beta}\Fc{}) \geq \frac{6 \Fs^2}{\text{SNR}
  \cdot \N_\text{sync}(\N_\text{sync}^2 - 1)(2\pi)^2}
\end{align}
Based on this expression, the constraint on $\epsilon$ can be approximated as
\begin{align}
  \label{eq:epsilon_approx}
  \epsilon \approx \frac{N}{2\pi} \sqrt{\frac{6}{\text{SNR} \cdot
  N_\text{sync}^3}} < 0.02
\end{align}
Designers will wish to minimize $N_\text{sync}$, since deterministic samples
devoted to synchronization do not carry information.  Suppose
$N_\text{sync} = 2^{10}$ and SNR = 10 dB.  Then $N$ must satisfy $N < 5316$ to
ensure $\epsilon < 0.02$.

Another practical constraint on $N$ is that it must be a power of two for
efficient IFFT and FFT operations at the transmitter and receiver.  No OFDM
waveform of which we are aware deviates from this norm.

Combining the power-of-two constraint with reasonable values of $N$ satisfying
(\ref{eq:epsilon_approx}), one can construct $\mathcal{S}$ as
\begin{align}
  \label{eq:set_S}
  \mathcal{S} = \left\{2^q : q \in \mathbb{N}, ~9 \leq q \leq 12\right\} 
\end{align}

The development leading to (\ref{eq:est_N}) assumed that
$\Fh \approx \Fr = \Fs$.  But of course, in the context of blind identification
of operational OFDM signals, the relationship of the receiver's sampling rate
$\Fr$ to $\Fs$ is unknown \emph{a priori}.  As will be revealed, the key to
accurate estimation of both $N$ and $\Fs$ is the power-of-two constraint on
$N$.

Let $\bar{\Fs}$ be a guess of $\Fs$ obtained by inspection of the power spectrum
of $y(n)$. This can be accomplished by a single wideband capture or by a sweep
of overlapping narrowband captures that collectively span a whole channel.  Note
that, besides $\bar{\Fs}$, one may also obtain from this inspection a guess of
the channel center frequency $\Fc{}$.  Bear in mind that even at high SNR it is
not possible to exactly determine $\Fs$ from the power spectrum because
subcarriers near the boundaries of an OFDM channel may be left unmodulated to
provide a frequency guard interval \cite{ancora2011ofdma}.  Let
$\Nr = \lfloor N \Fr/\Fs \rceil$ be the approximate number of receiver samples
in the useful symbol interval $T = N/\Fs$, where $\lfloor \cdot \rceil$ denotes
rounding to the nearest integer.  Also let $\eta = \Fr/\bar{\Fs}$ be the
estimated sampling rate ratio, and suppose that $|\bar{\Fs} - \Fs|/\Fs < p$ for
some $0 < p \ll 1$.  Then for each $b \in \mathcal{S}$, a set of corresponding
values of $\Nr$ can be constructed that accounts for the uncertainty in
$\bar{\Fs}$:
\begin{align}
  \label{eq:set_Srb}
  \mathcal{S}_{\text{r}b} = \left\{ \tau \in \mathbb{N} : 
  b \eta (1-p) \leq \tau \leq b \eta (1+p)\right\}
\end{align}
The full set of possible values of $\Nr$ is the union of these:
\begin{align}
  \label{eq:set_Sr}
  \mathcal{S}_{\text{r}}  = \bigcup_{b \in \mathcal{S}} \mathcal{S}_{\text{r}b} 
\end{align}
In other words, for every $b \in \mathcal{S}$, $\mathcal{S}_{\text{r}}$ contains
an interval of corresponding possible values of $\Nr$ whose width depends on the
assumed accuracy of $\bar{\Fs}$.  For convenience, define
$f_\text{r} : \mathcal{S}_\text{r} \rightarrow \mathcal{S}$ as the function that
maps possible values in $\mathcal{S}_\text{r}$ to the corresponding value in
$\mathcal{S}$; i.e., $\forall \tau \in \mathcal{S}_{\text{r}b}, f_\text{r}(\tau) = b$.

\begin{figure}[t]
\centering
\includegraphics[width=9cm]
{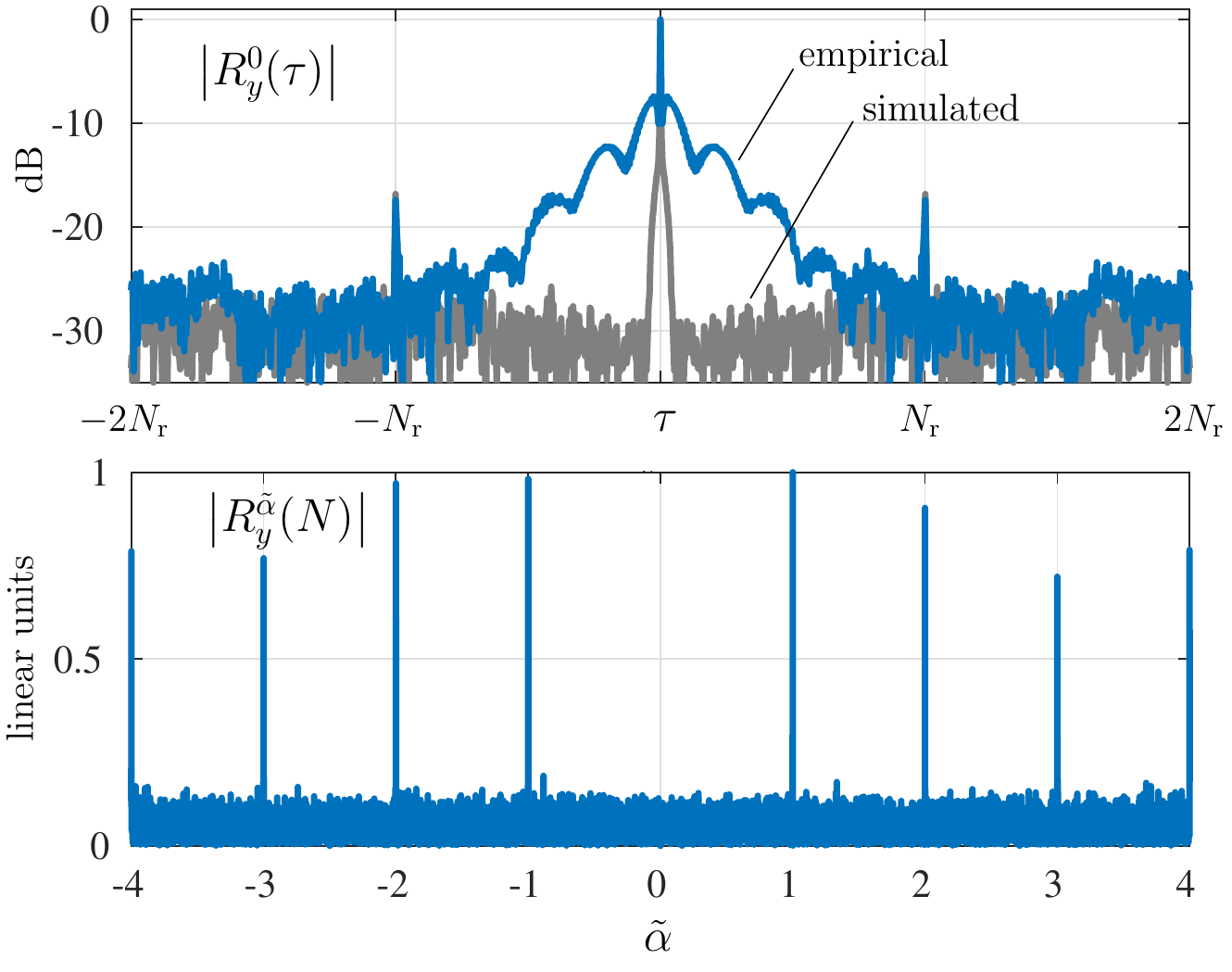}
\caption{Top: Cyclic autocorrelation function at $\alpha = 0$ for an empirical
  Starlink signal with $\text{SNR} = 5.5$ dB captured through the narrowband
  pipeline with capture interval approximately centered on an $\bar{\Fs}$-wide
  OFDM channel (blue), and for a simulated OFDM signal with iid Gaussian 4QAM
  symbols (gray).  The simulated signal has been Doppler-adjusted, passed
  through a simulated AWGN channel, lowpass filtered, and resampled at 62.5 MHz
  to match the empirical signal's Doppler, SNR, bandwidth, and sampling
  rate. The total number of samples $M$ used to estimate $R_y^\alpha(\tau)$ via
  (\ref{eq:cyclic_autocorr_empirical}) amounts to 10 ms of samples at
  $\Fr = 62.5$ MHz, which turn out to span just over 7 frames.  Bottom: Cyclic
  autocorrelation as a function of the normalized frequency
  $\tilde{\alpha} = \alpha(\N + \Ng)$ for
  $\alpha \in \left\{p/(N + \Ng) : p \in \mathbb{R}\right\}$, derived from the
  same empirical data as the blue trace in the top panel but resampled at
  ${\Fs} = 240$ MHz.  The peak at the fundamental cyclic frequency corresponding
  to the period $\Ng + \N$ appears at $\tilde{\alpha} = 1$; other peaks appear
  at harmonics of this fundamental.}
\label{fig:paramsPlotNandNgP}
\end{figure}

With these preliminaries we may recast the estimator in (\ref{eq:est_N}) for the
case in which $\Fs$ is only approximately known and may be significantly
different from $\Fr$:
\begin{align}
  \label{eq:N_est_operational}
   \hat{N} = f_\text{r}\left(\argmax_{\tau \in \mathcal{S}_\text{r}} \left|R^0_y(\tau)\right| \right)
\end{align}
Here, $R^0_y(\tau)$ is calculated by (\ref{eq:cyclic_autocorr_empirical}).  This
estimator works well for simulated OFDM signals, but must be augmented with a
validation step when applied to operational signals due to the phenomenon
manifest in Fig. \ref{fig:paramsPlotNandNgP}.  The blue trace in the top panel
shows that captured Starlink data exhibit a clear peak in
$\left|R^0_y(\tau)\right|$ at $\tau = \Nr$.  But the peak's magnitude is less
than that at other plausible values $\tau \in \mathcal{S}_\text{r}$ due to a
prominent central lobe in the empirical cyclic autocorrelation function.  This
lobe is due to a slower autocorrelation rolloff with increasing $|\tau|$ as
compared to a simulated OFDM signal with equivalent $\Nr$, $\beta$, SNR, $\Fh$,
and $\Fr$ (gray trace).  The slow rolloff indicates significant redundancy in
the received signal $y(n)$ at short offsets.  Such redundancy doubtless stems
from some combination of (i) strong error correction coding, (ii) inherent
redundancy in the data stream owing to light or negligible data compression in
an effort to achieve low latency, and (iii) adjacent-OFDM-symbol correlation
caused by pilot symbols.  The regular scalloped profile of the rolloff suggests
that (i) and (iii) may be the most important factors.

In any case, to prevent the maximization in (\ref{eq:N_est_operational}) from
choosing a value of $\tau$ at which $\left|R^0_y(\tau)\right|$ is large only
because of the prominent central autocorrelation lobe, $\hat{N}$ is accepted as
valid only if
\begin{align}
  \label{eq:validation_step}
  \frac{\max_{\tau \in \mathcal{S}_{\text{r}b}}
  \left|R^0_y(\tau)\right|}{\min_{\tau \in \mathcal{S}_{\text{r}b}}
  \left|R^0_y(\tau)\right|} > \nu, \quad b = \hat{N}
\end{align}
for some threshold $\nu$.  The point of this test is to ensure that the peak
value is sufficiently distinguished from others in its neighborhood, a condition
that does not hold within the wide central lobe of $\left|R^0_y(\tau)\right|$.
If this validation step fails, then $\mathcal{S}$ is redefined as
$\mathcal{S} \leftarrow \mathcal{S} \setminus \hat{N}$ and
(\ref{eq:N_est_operational}) is applied again, etc.  Empirically, we find that
for Starlink Ku-band downlink signals a threshold value $\nu = 10$ dB is
adequate to ensure that spurious maxima are excluded.  Note that one must choose
$p$ sufficiently large to ensure exploration of off-peak values in the
validation test.  This is especially important when $\Fh$ is significantly
smaller than $\Fr$, in which case the peak at $\left|R^0_y(\Nr)\right|$ may be
several samples wide.

\subsection{Estimation of  \normalfont  $\Fs$}
Having obtained $\hat{N}$, it is straightforward to obtain a more accurate
estimate of $\Fs$.  For $b = \hat{N}$, define
\begin{align}
  \label{eq:Nr_est_operational}
   \hat{\Nr} = \argmax_{\tau \in \mathcal{S}_{\text{r}b}} \left|R^0_y(\tau)\right|
\end{align}
Note that $\hat{\Nr}/\hat{N} \approx \Fr/\Fs$ and that, owing to the way blocks
of bandwidth are allocated by regulatory agencies, $\Fs$ is extremely likely to
be an integer multiple of 1 MHz.  Therefore, for $\Fr$ and $\Fs$ expressed in
MHz, an estimator for $\Fs$ is given by
\begin{align}
  \label{eq:Fs_est}
  \hat{\Fs} = \left\lfloor \frac{\hat{N}\Fr}{\hat{\Nr}} \right\rceil
\end{align}
The key to this estimator's accuracy is the power-of-two constraint on
$\hat{N}$.

\subsection{Resampling}
\label{sec:resampling}
Estimation of the remaining OFDM parameters of interest is facilitated by
resampling $y(n)$ at $\hat{\Fs}$.  Recall from (\ref{eq:received signal model})
that $y(n)$ is natively sampled at $\Fr$ after lowpass filtering with bandwidth
$\Fh$.  For the narrowband capture mode, resampling at $\hat{\Fs}$ implies a
sampling rate increase, which can be modeled as \cite{proakis2008digital}
\begin{align}
  \label{eq:resampling_y}
  y_\text{r}(m) = \sum_{n \in \mathbb{Z}} y(n) \text{sinc}(m\Fr/\hat{\Fs} - n) 
\end{align}
where $\sinc(x) = \sin(\pi x)/(\pi x)$.  Note that the useful frequency content
of the signal, $|f| < \Fh$, remains unchanged.  For the wideband capture mode,
resampling at $\hat{\Fs}$ implies conversion to a lower sampling rate after
lowpass filtering with a new lower $\Fh$. For notational simplicity, in what
follows we will drop the subscript from $y_\text{r}$.  Thus, $y(n)$ will
hereafter denote the received signal with bandwidth $\Fh$ (possibly less than
the original) and sampling rate $\hat{\Fs}$.

\subsection{Estimation of  \normalfont  $\Ng$}
\label{sec:estim-norm-ng}
Estimation of $\Ng$ begins by constructing the set $\mathcal{S}_\text{g}$ of
possible values of $N + \Ng$.  As with $\mathcal{S}$, this is informed by design
constraints.  From (\ref{eq:ofdm_throughput}) it is clear that signal designers
will wish to minimize $\Ng$, but this is subject to the constraint that
$\Tg = \Ng/\Fs$ exceeds the channel's delay spread under all but the most
extreme operating conditions.  Worst-case 95\% root-mean-square delay spread for
the Ku-band was found in \cite{cid2015wideband} to be $\Td = 108$ ns.
Conservatively considering a range of values from half to twice this amount, and
assuming that, for ease of implementation, $\Ng$ is even, one can construct
$\mathcal{S}_\text{g}$ as
\begin{align}
  \label{eq:Sg_construction}
\mathcal{S}_\text{g} = \left\{2q + b : b \in \mathcal{S},~q \in \mathbb{N},
 ~ \Td\hat{\Fs}/4 \leq q \leq \Td\hat{\Fs} \right\}
\end{align}
With $y(n)$ sampled at $\hat{\Fs}$, estimation of $\Ng$ then proceeds as in
(\ref{eq:est_Ng}) except that $R_y^\alpha(\hat{N})$ is calculated via
(\ref{eq:cyclic_autocorr_empirical}) and $\mathcal{A}(\xi)$ is reduced to the
finite set
$\mathcal{A}(\xi) = \left\{ p/\xi : p \in \mathbb{Z}, |p| \leq N_p \right\}$.
for some finite $N_p$.

The accuracy of this estimator as a function of $N_p$ is analyzed in
\cite{bouzegzi2010new}, where it is shown that no improvement attains to values
of $N_p$ above $N/\Ng$.  In practice, when applied to Starlink signals captured
via the narrowband mode, estimator performance was reliable for $N_p$ as low as
1 provided that the number of samples $M$ in
(\ref{eq:cyclic_autocorr_empirical}) covered at least one frame
($M \geq \Tf \Fs$) and that $\text{SNR} > 3.5$ dB.

The lower panel in Fig. \ref{fig:paramsPlotNandNgP} shows a version of
$|R_y^\alpha(N)|$ from empirical Starlink data at $\text{SNR} = 5.5$ dB that has
been normalized so that the cyclic frequencies are integers.  The span of cyclic
frequencies shown corresponds to $N_p = 4$.

\subsection{Estimation of \normalfont $\Tf$}
Each frame contains one or more OFDM symbols with predictable elements, called
synchronization sequences, that enable receiver time and frequency
synchronization.  A peak emerges in the cyclic autocorrelation $R^0_y(\tau)$
when one synchronization sequence is aligned with its counterpart from a nearby
frame.  Thus, estimation of the frame period $\Tf$ is also based on
$R^0_y(\tau)$ as calculated in (\ref{eq:cyclic_autocorr_empirical}), but now
with the number of samples $M$ large enough to cover multiple adjacent frames.

Let $\Nf = \Tf \Fs$ be the frame period expressed in number of samples, and let
$\mathcal{S}_\text{f}$ be the set of possible values of $\Nf$.  By inspection of
the empirical signal spectrogram during a period of sparse traffic, one can
easily obtain an upper bound $\Tm$ on the smallest active signal interval.
Observe that this may be a loose upper bound on $\Tf$ because the smallest
active interval observed may actually be multiple frames.  One can then
construct a conservative $\mathcal{S}_\text{f}$ as follows:
\begin{align}
  \label{eq:Sf}
  \mathcal{S}_\text{f} = \left\{ q \in \mathbb{N} : \hat{N} + \hat{\Ng} < q \leq
  \hat{\Fs} \Tm \right\}
\end{align}

Considerations of expected signal numerology are once again useful in the case
of estimating $\Tf$.  While $\Tf$ need not be an integer number of milliseconds,
$\Nf$ is likely to be an integer for ease of signal generation, and, more
importantly, the frame rate $\Ff = 1/\Tf$ is almost certainly integer
number of Hz for ease of frame scheduling across the constellation.  Therefore,
for $\hat{\Fs}$ expressed in Hz, an effective estimator for $\Tf$ is given by
\begin{align}
  \label{eq:Tf_estimator}
  \hat{\Tf} = \left\lfloor \hat{\Fs} \left(\argmax_{\tau \in
  \mathcal{S}_\text{f}}\left| R_y^0(\tau)\right| \right)^{-1} \right\rceil^{-1}
\end{align}
Note that, as for the estimators of $N$ and $\Ng$, this estimator for $\Tf$ is
robust to nonzero Doppler provided that $\beta \Fc{} \Tss \ll 1$, where
$\Tss$ is the longest time interval of any contiguous synchronization
sequence.

\subsection{Symbol and Carrier Frequency Synchronization}
\label{sec:symb-carr-freq}
Estimating the remaining parameters in Table \ref{tab:parameters-description}
and any synchronization sequences requires both OFDM symbol synchronization and
carrier frequency synchronization.  Such synchronization must be carried out
blindly, since the very sequences designed to enable it are unknown.  

Let $n_{mik}$ be the index of the $k$th sample in the $i$th OFDM symbol of the
$m$th frame, assuming zero-based indexing of $k$, $i$, and $m$.  For some
$m,i \in \mathbb{N}$ with $i < \Nsf$, we wish to find $n_{mi0}$
and the value of the CFO parameter $\beta$ that applies at $n_{mi0}$, denoted
$\beta_{mi}$.

When frame traffic is low enough that gaps are present between frames, it is
possible to observe an abrupt increase in sample energy $|y(n)|^2$ at the
beginning of a frame, which allows approximation of $n_{m00}$, the index of the
first sample in the first OFDM symbol of the frame.  By adding integer multiples
of $\hat{N} + \hat{\Ng}$, one can then approximate $n_{mi0}$ for all
$i \in (0,\Nsf)$.  Let $\bar{n}_{mi0}$ be an approximate value for $n_{mi0}$.
Then $\mathcal{S}_{mi}$, the set of possible values of $n_{mi0}$, may be
constructed as
\begin{align}
  \label{eq:Sn}
  \mathcal{S}_{mi} = \left\{ n \in \mathbb{Z} : \left| n - \bar{n}_{mi0}\right|
  \leq d \right\}
\end{align}
with $d$ large enough to account for uncertainty in $\bar{n}_{mi0}$.

Let $\mathcal{B}_{mi}$ be the set of possible values of $\beta_{mi}$.  One might
think that the range of \emph{a priori} uncertainty in $\beta_{mi}$ is small
because, for known receiver location and time, and known transmitting satellite
ephemeris, the line of sight velocity $\vlos$ can be readily calculated, from
which $\beta$ can be calculated as $\beta = \vlos/c$.  But recall from Section
\ref{sec:rece-sign-model} that $\beta$ also accounts for any frequency offset
between the transmitter and receiver oscillators.  In the present context, such
an offset may arise not only because of disagreement between the oscillators,
but also due to uncertain knowledge of $\Fc{}$, the center frequency of the OFDM
channel captured to produce $y(n)$.  As a consequence, the range of $\beta_{mi}$
values included in $\mathcal{B}_{mi}$ may be several times larger than what
would be predicted based on $\vlos/c$ alone.  Let $\bar{\beta}_{mi}$ be a prior
estimate of $\beta_{mi}$ based on ephemeris calculations and any other relevant
prior information, $\beta_\text{m}$ be the maximum offset from
$\bar{\beta}_{mi}$ considered, and
$\Delta \beta = \epsilon \hat{\Fs}/\hat{N}{\bar{\Fc{}}}$ be the search stride,
chosen to satisfy (\ref{eq:beta_tilde_reqt}), where $\bar{\Fc{}}$ is both an
\emph{a priori} estimate of $\Fc{}$ obtained by inspection of the power spectrum
of $y(n)$, and the exact center of the band captured to produce $y(n)$.  Then
$\mathcal{B}_{mi}$ may be constructed as
\begin{align}
  \mathcal{B}_{mi} = \left\{q \Delta \beta : q \in \mathbb{Z}, ~ \left|q \Delta
  \beta - \bar{\beta}_{mi}\right| \leq \beta_\text{m} \right\}
\end{align}

\begin{figure}[t]
\centering
\includegraphics[width=9cm]
{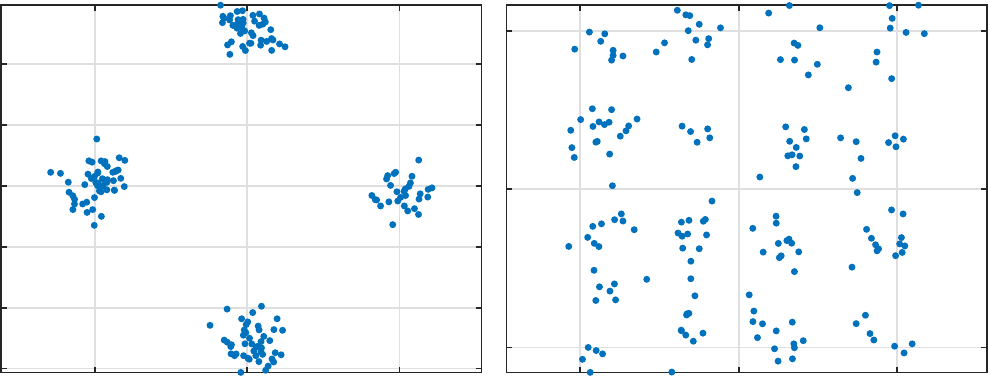}
\caption{Empirical Starlink symbol constellations for 4QAM (left) and 16QAM
  (right) OFDM modulation.}
\label{fig:constellationPlot}
\end{figure}

By a simultaneous search through the values in $\mathcal{S}_{mi}$ and
$\mathcal{B}_{mi}$, one may estimate $n_{mi0}$ and $\beta_{mi0}$ with sufficient
accuracy to enable standard receiver processing of each corresponding OFDM
symbol in $y(n)$, leading to recovery of the relevant original information
symbols $\{X_{mik}\}$.  Fig. \ref{fig:constellationPlot} shows the successful
result for a portion of one frequency-domain OFDM symbol with 4QAM modulation
and another symbol with 16QAM modulation. Tight constellation clusters like
those in the left panel only emerge when SNR is sufficiently high (15 dB in this
case) and when the estimates of $n_{mi0}$ and $\beta_{mi}$ are accurate enough
that ISI and ICI are negligible.  Otherwise, the clusters become elongated (due
to mild ISI or ICI), or they experience a complete collapse toward the origin
(severe lack of synchronization).  Clearly then, symbol constellations can be
used to develop a score function that increases with synchronization accuracy.

Let
$\text{SC} : \mathcal{S}_{mi} \times \mathcal{B}_{mi} \rightarrow
\mathbb{R}_{+}$ be such a function, with trial synchronization values
$n \in \mathcal{S}_{mi}$ and $\beta \in \mathcal{B}_{mi}$ as arguments.
Algorithm~\ref{algo:SC} shows the computations underlying $\text{SC}$.  First,
an OFDM-symbol-length block of samples is isolated starting at the trial index
$n$.  The block is then resampled and frequency shifted to undo the effects of
nonzero $\beta$, after which its cyclic prefix is discarded and the remaining
samples are converted to the frequency domain via an FFT.  The resulting
received information symbols in $\vb{Y}$ cluster as shown by the examples in
Fig. \ref{fig:constellationPlot}.  Assuming $b_\text{s}$ bits per symbol,
$2^{b_\text{s}}$ clusters will be present.  These are identified automatically
via $k$-means clustering.  For $b_\text{s} \leq 2$, the function's output $s$ is
the empirical SNR, calculated as the mean of the squared magnitude of each
cluster's centroid divided by twice the cluster's variance.  Note that $s$ is
insensitive to rotations of the constellation due to the unknown reference phase
of the symbols $\{X_{mik}\}$.

\begin{algorithm}
\SetKwInOut{Input}{Input}
\SetKwInOut{Output}{Output}
\Input{$n \in \mathcal{S}_{mi}, ~\beta \in \mathcal{B}_{mi}$}
\Output{$s \in \mathbb{R}_{+}$}

$\vb{y} = [y(n), y(n+1), \dots, y(n+\hat{N} + \hat{\Ng} - 1)]$

$\vb{t}_y = [0:\hat{N} + \hat{\Ng} - 1]/\hat{\Fs}$

$[\vb{y}, \vb{t}_y] = \texttt{resample}(\vb{y}, \vb{t}_y, (1-\beta)\hat{\Fs})$

\For{$i = 0:\hat{N} + \hat{\Ng} - 1$}{
  $\vb{y}(i) = \vb{y}(i)\exp\left(j 2 \pi \beta \bar{\Fc{}} \vb{t}_y(i) \right)$
}

$\vb{y} = \vb{y}(\hat{\Ng} : \hat{\Ng} + \hat{\N} - 1)$

$\vb{Y} = \texttt{fft}(\vb{y})$

$[\vb{c},\vb{\sigma}] = \texttt{kmeans}(\vb{Y},2^{b_\text{s}})$

\For{$i = 0:2^{b_\text{s}}-1$}{
  $\vb{s}(i) = |\vb{c}(i)|^2/2\vb{\sigma}^2(i)$
}

$s = \texttt{mean}(\vb{s})$

\caption{${\text{SC}}(n,\beta)$}
\label{algo:SC}
\end{algorithm}
\noindent

With $\text{SC}$, construction of the estimator for ${n}_{mi0}$ and
${\beta}_{mi0}$ is straightforward:
\begin{align}
  \label{eq:SC_est}
  \hat{n}_{mi0},~\hat{\beta}_{mi} = \argmax_{\footnotesize \ba{c} n \in \mathcal{S}_{mi}\\ \beta \in
  \mathcal{B}_{mi} \ea} \text{SC}(n,\beta)
\end{align}

This estimator was found to work well on both types of standard OFDM symbol
modulation found in the captured Starlink signal frames, namely 4QAM
($b_\text{s} = 2$) and 16QAM ($b_\text{s} = 4$), even when the signals' SNR was
too low to ensure error-free cluster identification, as in the right panel in
Fig. \ref{fig:constellationPlot}.  But the estimator failed unexpectedly when
applied to the first OFDM symbol interval in each frame.  Closer examination
revealed that this interval does not contain an OFDM symbol but rather a
repeating pseudorandom time-domain sequence.  Nonetheless, estimates of
$n_{m00}$ and $\beta_{m00}$ were accurately obtained as
$\hat{n}_{m00} = \hat{n}_{m10} - \hat{N} - \hat{\Ng}$ and
$\hat{\beta}_{m0} = \hat{\beta}_{m1}$.

\subsection{Estimation of the Synchronization Sequences}
\label{sec:estim-synchr-sequ}
Estimating the synchronization sequences embedded in each Starlink frame is one
of this paper's key contributions.  To this end, one must first locate the
sequences, i.e., determine which OFDM symbol intervals within a frame contain
predictable features.  Recall that, by definition, synchronization sequences are
predictable from the perspective of the user terminal.  For public-access OFDM
signals such as Wi-Fi, WiMAX, LTE, etc., they are not only predictable but
constant from frame to frame.  Presuming the same for Starlink signals, locating
such sequences within a frame is a matter of isolating individual OFDM symbol
intervals and correlating these across multiple frames to determine whether the
candidate intervals contain features that repeat from frame to frame.  Isolating
OFDM symbol intervals is possible at this stage because $\hat{n}_{mi0}$,
$\hat{N}$, and $\hat{\Ng}$, and are available.

This procedure revealed that the first OFDM symbol in each Starlink frame, the
one starting at sample index $n_{m00}$, contains a synchronization sequence.
The interval was found to lack any discernible constellation structure when
viewed in the frequency domain.  But its cross-correlation against first symbol
intervals in neighboring frames revealed a pattern of peaks indicating that the
interval is composed of 8 repetitions of a time-domain-rendered subsequence of
symbols of length $\hat{N}/8$, with the first instance inverted. (A complete
model of this synchronization sequence is presented in a later section.)
Estimation of the exact symbol values was only possible using data obtained via
the wideband capture mode, since the subsequence's frequency content spans the
whole of $\Fs$.  Despite the low SNR of the wideband capture mode, knowledge of
$\hat{\beta}_{m0}$, $\hat{n}_{mi0}$, $\hat{N}$, and $\hat{\Ng}$ allowed the 8
subsequence repetitions to be stacked and summed coherently to reveal the unique
subsequence values, which will be presented in a following section.  The 8
subsequence repetitions are prepended by a cyclic prefix of length $\hat{\Ng}$.
Borrowing language from the LTE specification, we call the full
$(\hat{N} + \hat{\Ng})$-length sequence the primary synchronization sequence
(PSS).  It was found that the PSS is not only identical across all frames from
the same Starlink satellite, but also identical across all satellites in the
constellation.

The second OFDM symbol interval, which starts at sample index $n_{m10}$, was
also found to contain a $(\hat{N} + \hat{\Ng})$-length synchronization sequence,
which we call the secondary synchronization sequence (SSS).  Unlike the PSS, the
SSS was found to be a standard OFDM symbol, with 4QAM modulation.  Estimating
the information symbols $\{X_{m1k}\}_{k=0}^{N-1}$ was possible even with
narrowband-mode-captured data because the received symbols that fell within the
narrowband mode's bandwidth were clearly observable (to within a phase offset),
as shown in the left panel of Fig. \ref{fig:constellationPlot}.  In other words,
with the high-SNR narrowband data, those elements of $\vb{Y}$ in
Algorithm~\ref{algo:SC} corresponding to frequencies within the $62.5$-MHz
narrowband window could be confidently assigned to one of four clusters.  At
this stage, it was not known whether the SSS was anchored with an absolute
initial phase so that the symbols $\{X_{m1k}\}_{k=0}^{N-1}$ would be constant
across $m$, or differentially encoded so that only $X_{m1(k+1)}^*X_{m1k}$ would
be constant, for $k \in [0,N - 2]$ . Moreover, the estimates $\hat{n}_{m10}$ for
various $m$ were not precise enough at this stage to ensure that corresponding
constellation clusters could be associated with each other from frame to frame.
Therefore, only differential values were initially estimated, with
$Y_{m1(k+1)}^*Y_{m1k}$ being an estimate of $X_{m1(k+1)}^*X_{m1k}$, where
$Y_{m1k}$ is the $k$th element of $\vb{Y}$ in Algorithm~\ref{algo:SC} for OFDM
symbol $i =1$ of frame $m$.

By successively shifting the $62.5$-MHz capture band across an OFDM channel of
width $\Fs$ in repeated captures, and by ensuring sufficient frequency overlap,
it was possible to confidently estimate each $X_{m1(k+1)}^*X_{m1k}$ such that
the full sequence $\{X_{m1k}\}_{k=0}^{N-1}$ could be determined to within two
unknown symbols, $X_{m12}$ and $X_{m1(N/2)}$.  The first of these is
unobservable from the differential estimates due to the presence of a
mid-channel ``gutter'' in which $X_{mik} = 0$ for $k\in \{0,1,N-2,N-1\}$; the
second is unobservable because it lies at the bottom edge of the frequency band.
By searching through all possible combinations of these two unknown symbols,
re-generating for each trial combination a candidate time-domain OFDM SSS
(prepended by the appropriate cyclic prefix), concatenating this candidate SSS
with the known time-domain PSS, and maximizing correlation against the first two
OFDM symbol intervals in received data frames, all while resampling and
frequency shifting the received data to account for nonzero $\beta$ as in
Algorithm~\ref{algo:SC}, it was possible to estimate $X_{m12}$ and $X_{m1(N/2)}$
and thereby completely determine the SSS.  As with the PSS, it was found that
the SSS is identical across all satellites in the Starlink constellation.

The last nonzero OFDM symbol in each frame, the one starting at sample index
$n_{mi0}$ with $i=301$, was also found to contain a
$(\hat{N} + \hat{\Ng})$-length synchronization sequence, which we call the coda
synchronization sequence (CSS).  Like the SSS, the CSS is a standard 4QAM OFDM
symbol whose information symbols $\{X_{mik}\}_{i = 301, k=0}^{k = N-1}$ can be
determined by inspection.  The CSS symbol constellation is rotated by 90 degrees
with respect to the SSS: whereas the SSS exhibits the diamond configuration
shown in the left panel of Fig. \ref{fig:constellationPlot}, the CSS's
constellation clusters form a box aligned with the horizontal and vertical axes.

The penultimate nonzero OFDM symbol in each frame, the one starting at sample
index $n_{mi0}$ with $i=300$, was found to contain some information symbols that
are constant from frame to frame.  But, unlike the SSS and the CSS, not all the
information symbols are constant.  We call the predictable elements of this
symbol the coda-minus-one synchronization sequence (CM1SS).

\subsection{Estimation of \normalfont $\Nsf$, $\Ndataframe$, {\it and} $\Tfg$}
Equipped with $\hat{n}_{mi0}$, $\hat{N}$, $\hat{\Ng}$, $\hat{\Tf}$, and
knowledge that the first two OFDM symbol intervals in each frame are
synchronization sequences, it is trivial to estimate $\Nsf$, $\Ndataframe$, and
$\Tfg$.  The estimated OFDM symbol duration is
$\hTsym = (\hat{N} + \hat{\Ng})/\hat{\Fs}$, and thus the estimated number of
whole symbol intervals in one frame is $\lfloor \hat{\Tf}/\hTsym \rfloor$, where
$\lfloor \cdot \rfloor$ denotes the floor function.  The final interval was
found to be vacant.  Thus, the estimated number of non-zero symbols in a frame
is
\begin{align}
  \hNsf = \lfloor \hat{\Tf}/\hTsym \rfloor - 1
\end{align}
Counting the PSS, SSS, CM1SS, and CSS as synchronization symbols, the
estimated number of non-synchronization symbols in a frame is
\begin{align}
  \hNdataframe = \hNsf - 4
\end{align}
Finally, the estimated frame guard interval---the vacant interval between
successive frames---is
\begin{align}
  \hTfg = \hat{\Tf} - \hNsf \hTsym
\end{align}

\subsection{Estimation of  \normalfont $\Fc{i}$}
Estimation of $\Fc{i}$, the center frequency of the $i$th Starlink OFDM channel,
is complicated by the exponential in (\ref{eq:received_with_dopp}) being a
function of both $\beta$ and the offset $\Fc{} - \bar{\Fc{}}$.  This implies
that an error in the \emph{a priori} estimate $\bar{F}_{\text{c}i}$ results in a
frequency offset just as with nonzero $\beta$.  But the two effects can be
distinguished by recognizing that compression or dilation of the modulation
$x(t)$ in (\ref{eq:received_with_dopp}) is solely a function of $\beta$.
Therefore, determination of $\Fc{i}$ begins by estimating the $\beta$ that
applies for the $i$th channel as expressed via $x(t)$, which may be done by
measuring a sequence of frame arrival times.

Assume that the local receiver clock used for downmixing and sampling the
received signal is short-term stable and GPS-disciplined, as with the 10-MHz
OCXO in Fig. \ref{fig:sig-cap}, so that it may be considered a true time
reference.  Let $\{\hat{n}_{m00}\}_{m \in \mathcal{M}}$ be the estimated indices
of samples that begin a frame for channel $i$, as determined by
(\ref{eq:SC_est}) or by correlation against the known PSS and/or SSS.  Note that
the set of frame indices in $\mathcal{M}$ may not have a regular spacing.  Let
the nominal time $t(m)$ and the received time $t_\text{r}(m)$ of frame
$m \in \mathcal{M}$ be
\begin{align*}
  t(m) = m \hat{\Tf}, \quad t_\text{r}(m) = \hat{n}_{m00}/\hat{\Fs}
\end{align*}
For intervals up to one second, which a study of frame timing revealed as the
cadence at which clock corrections are applied onboard the Starlink satellites,
the relationship between $t(m)$ and $t_\text{r}(m)$ can be accurately modeled as
a second-order polynomial
\begin{align*}
  t_\text{r}(m) = a_0 + a_1\left(t(m) - t(m_0)\right) + a_2\left(t(m) -
  t(m_0)\right)^2
\end{align*}
where $m_0 = \min \mathcal{M}$.  Let $\{\hat{a}_i\}_{i = 0}^2$ be coefficient
estimates obtained via least squares batch estimation.  Then
$\bar{\beta}_{m_00} = \hat{a}_1$ is the modulation-estimated $\beta$ value that
applies at the beginning of frame $m_0$.  Let $\hat{\beta}_{m_00}$ be the value
of $\beta$ that applies at the same instant, as estimated by (\ref{eq:SC_est}).
Also, recall that $\bar{F}_{\text{c}{i}}$ is both the \emph{a priori} estimate
of $\Fc{i}$ assumed in (\ref{eq:SC_est}) and the exact center of the band
captured to produce the $y(n)$.  Then
\begin{align}
  \label{eq:Fci_est}
  \hat{F}_{\text{c}i} = \left\lfloor \frac{\bar{F}_{\text{c}{i}}}{1 +
  \bar{\beta}_{m_00} - \hat{\beta}_{m_00}} \right\rceil
\end{align}
is an estimator of $\Fc{i}$, where $\hat{F}_{\text{c}i}$ and
$\bar{F}_{\text{c}{i}}$ are expressed in MHz.  Rounding to the nearest MHz is
justified for the same reasons given in connection with (\ref{eq:Fs_est}).

\section{Results}
\label{sec:results-sign-ident}
\begin{figure}[t]
  \centering
  \includegraphics[width=0.5\textwidth]{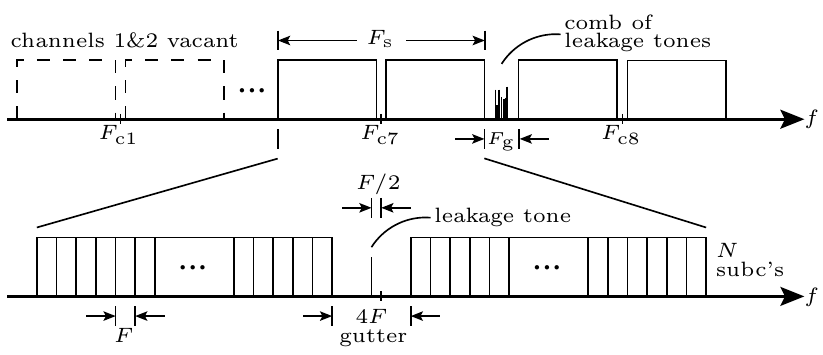}
  \caption{Channel layout for the Ku-band Starlink downlink.}
  \label{fig:channel-diagram2}
\end{figure}

\begin{figure*}[t]
  \centering
  \includegraphics[width=0.75\textwidth]{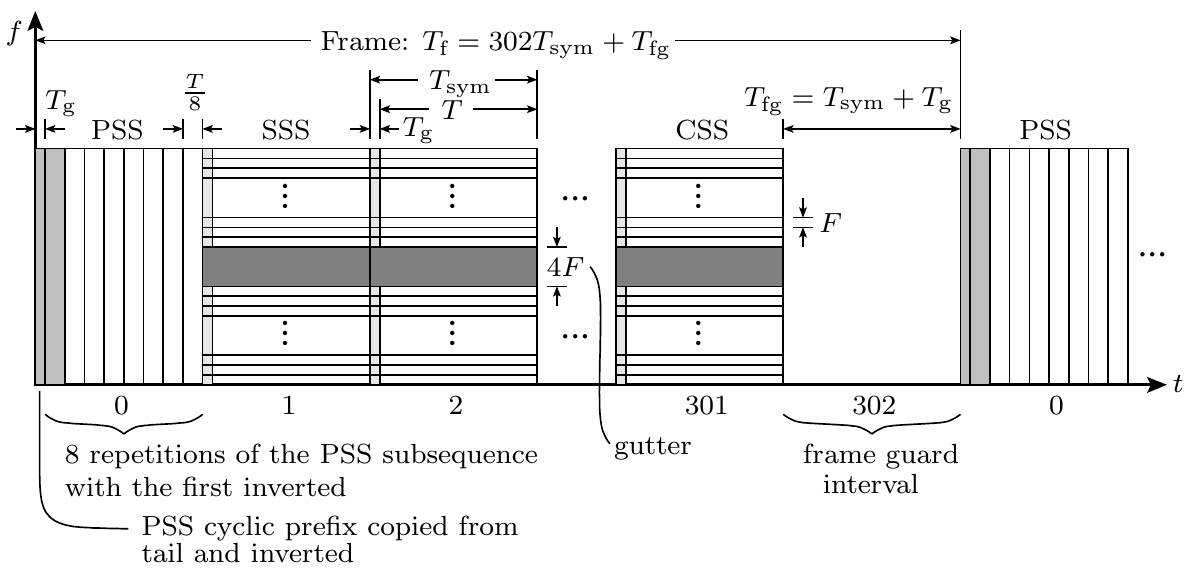}
  \caption{Frame layout for the Ku-band Starlink downlink along time-frequency
    dimensions.}
  \label{fig:frame-diagram}
\end{figure*}

Application of the foregoing blind signal identification procedure yields the
parameter values given in Table \ref{tab:parameter-values} for the Starlink
Ku-band downlink.  Figs. \ref{fig:channel-diagram2} and \ref{fig:frame-diagram}
offer graphical representations of the channel and frame layouts. The PSS was
found to be composed of eight repetitions of a length-$N/8$ subsequence
prepended by a cyclic prefix.  As shown in Fig. \ref{fig:frame-diagram}, the
cyclic prefix and the first instance of the repeated subsequence have inverted
polarity relative to the remainder of the PSS. The time-domain expression of the
PSS can be written as
\begin{align}
  x_{m0}(t) & = \sum_{k=-\Ng}^{\N-1} \sinc\left[t \Fs- k - \Ng\right] p_k
  \\
  \label{eq:pk_expression}
  p_k & = \exp\left(j\pi \left[
    \mathbf{1}_{\mathcal{P}}(k) - \frac{1}{4} -
    \frac{1}{2}\sum_{\ell=0}^{k\Mod\frac\N8} b_\ell
        \right]\right) \\
  \label{eq:bell_expression}
  b_\ell & =  2\left(\left\lfloor
                \frac{q_\text{pss}}{2^\ell}\right\rfloor \Mod 2\right)-1
\end{align}
where $\mathbf{1}_{\mathcal{P}}(k)$ is the indicator function, equal to unity
when $k \in \mathcal{P}$ and zero otherwise, and
$\mathcal{P} = \{k \in \mathbb{Z} : k < N/8\}$.  The indicator function rotates
the phase by $\pi$ for $k < N/8$ to invert the cyclic prefix and the first
repetition of the PSS subsequence.  The PSS subsequence
$(p_k)_{k = N/8}^{2N/8-1}$ is a symmetric differential phase shift keying
(symmetric DPSK) encoding of a length-$127$ maximal-length linear-feedback shift
register (LFSR) sequence (m-sequence). In this modulation, each bit of the
m-sequence indicates a positive or negative $\pi/2$ phase rotation. The
m-sequence can be generated using a 7-stage Fibonacci LFSR with primitive
polynomial $1 + D^3 + D^7$ and initial state
$(a_{-1},\dots,a_{-7}) = (0, 0, 1, 1, 0, 1, 0) $, following the convention in
\cite{dinan1998spreading}.  Suppose that the LFSR's output
$a_{0}, a_1, \dots, a_{126}$ is stored as a 127-bit number with $a_{0}$ as MSB
and $a_{126}$ as LSB. Appending this number with a $0$ yields the 128-bit
hexadecimal number that appears in (\ref{eq:bell_expression}):
\begin{align*}
q_\text{pss} =
  \parbox[c]{\textwidth}{\ttfamily\small
  C1B5\:D191\:024D\:3DC3\:F8EC\:52FA\:A16F\:3958}
\end{align*}
To ensure correct interpretation of (\ref{eq:pk_expression}) and
(\ref{eq:bell_expression}), we list the first 8 values of the PSS subsequence:
\begin{align*}
  & p_k = \exp \left(j\pi \left[1/4 + q_k/2 \right] \right), \quad k \in \{N/8, \dots, N/8 + 7\} \\
  & (q_{N/8}, \dots, q_{N/8 + 7}) = (0,1,2,1,0,1,0,1)
\end{align*}

\begin{table}[t]
  \centering
  \caption{Starlink Downlink Signal Parameter Values}
  \begin{tabular}[c]{lcl}
    \toprule
    Parameter &  Value & Units \\ \midrule
    \Fs        & 240 & MHz \\
    \N       & 1024 & \\
    \Ng        & 32 & \\
    \Tf    & 1/750 & s \\
    \Tfg       & $68/15 = 4.5\overline{33}$ & $\mu$s\\
    \Nsf & 302 & \\
    \Ndataframe& 298 & \\
    \T         & $64/15 = 4.2\overline{66}$  & $\mu$s  \\
    \Tg        & $2/15 = 0.1\overline{33}$ & $\mu$s \\
    \Tsym      & 4.4 & $\mu$s \\
    \F       & 234375  & Hz \\
    \Fc{i}& $10.7 + F/2 + 0.25(i-1/2)$ & GHz \\
    \DeltaFcenter   & 250 & MHz \\
    \Fg    & 10 & MHz \\
    \bottomrule
  \end{tabular}
  \label{tab:parameter-values}  
\end{table}

The time-domain expression of the SSS can be written as $x_{m1}(t)$ from
(\ref{eq:generic-symbol}) with the complex coefficients given by
\begin{align}
  X_{m1k} &= \left\{ \ba{ll} \exp (j\theta_k), & k \in \{2,\dots,N-3\} \\
  0, & \text{otherwise} \ea \right. \\
  \theta_k &= s_k \pi/2  \\
    \label{eq:sk}
     s_k & = \left\lfloor
                \frac{q_\text{sss}}{4^{k-2}}\right\rfloor \Mod 4  
\end{align}
where $q_\text{sss}$ is the hexadecimal number
\begin{align*}
 q_\text{sss} = & \parbox[r]{\textwidth}{\ttfamily\small
 ~~BD\:565D\:5064\:E9B3\:A949\:58F2\:8624\:DED5}\\
  & \parbox[r]{\textwidth}{\ttfamily\small
 6094\:6199\:F5B4\:0F0E\:4FB5\:EFCB\:473B\:4C24\\
 B2D1\:E0BD\:01A6\:A04D\:5017\:DE91\:A8EC\:C0DA\\
 09EB\:FE57\:F9F1\:B44C\:532F\:161C\:583A\:4249\\
 0A5C\:09F2\:A117\:F9A2\:8F9B\:2FD5\:47A7\:4C44\\
 BABB\:4BE8\:5DA6\:A62B\:1235\:E2AD\:084C\:0018\\
 0142\:A8F7\:F357\:DEC4\:F313\:16BC\:58FA\:4049\\
 09A3\:FCA7\:F88E\:4219\:02B6\:A258\:0AE8\:0308\\
 03F6\:5809\:DB34\:7F59\:0DBC\:46F0\:10EB\:E3A2\\
 5C06\:0D74\:429F\:C46B\:DF9B\:6371\:9279\:798D\\
 232C\:5ABA\:2741\:22FF\:66AD\:7E44\:9F44\:CB40\\
 C49C\:24A1\:E262\:9F5B\:FE82\:CE53\:1FDC\:34F8\\
 C64A\:43A9\:63F4\:0D5B\:71BD\:E6FB\:2F13\:492D\\
 6F2E\:8544\:B21D\:4497\:22C6\:3518\:0342\:CD00\\
 26A1\:E7F7\:E80E\:91B1\:75E8\:52F9\:1976\:7E5A\\
 F9B6\:E909\:AF36\:2F52\:18E2\:B908\:DC00\:5803}
\end{align*}
To ensure correct interpretation of (\ref{eq:sk}), we provide the first 8 values
of $s_k$ corresponding to nonzero $X_{m1k}$:
\begin{align*}
  (s_2,\dots,s_{9}) = (3, 0, 0, 0, 0, 2, 1, 1) 
\end{align*}

The CSS and the CM1SS will be presented in a later publication to provide
adequate space for a detailed presentation and discussion of their
characteristics.

\section{Discussion}
Our blind signal identification process reveals a Starlink Ku-band downlink
signal that is elegantly simple.  Unlike LTE and 5G New Radio (5G NR), whose
bandwidth and duplexing scheme may vary from region to region, and whose cyclic
prefix length may vary with time, Starlink employs fewer modes of operation.
This section offers observations on salient features of the Starlink signal.

\subsection{Channel Layout}
As shown in Fig. \ref{fig:channel-diagram2}, a total of eight channels, each
with a bandwidth of $\Fs = 240$ MHz, span the band allocated for Starlink's
Ku-band downlink. In principle, multiple channels could be active simultaneously
within a service cell.  We assume that neighboring cells are serviced with
different channels to avoid inter-cell interference, as described in
\cite{iannucci2022fusedLeo}, but we were not able to verify this with our
limited experimental setup.  The lower two channels, those centered at $\Fc{1}$
and $\Fc{2}$, are currently vacant.  This likely reflects a concession SpaceX
has made to avoid interfering with the 10.6-10.7 GHz radio astronomy band.

Each channel's central four subcarriers are vacant, leaving a mid-channel
gutter.  Reserving such a gutter is a common practice in OFDM; otherwise,
leakage from a receiver's mixing frequency may corrupt central information
symbols.  In Starlink's case, a transmitter-side leakage tone is present in some
gutters for some satellites.  For example, a leakage tone was found in the
gutter of channel 5 on the Starlink satellite with identifier 3262, channel 6 on
Starlink 3503, and channel 5 on Starlink 2409, whereas for other satellites no
leakage tones were observed for the same channels.  Interestingly, the $i$th
channel's center frequency, $\Fc{i}$, is $\F/2$ higher than the channel's
midpoint, which lies in the center of the mid-channel gutter.  A gutter leakage
tone, if present, resides at the channel midpoint.

A guard band with a generous bandwidth $\Fg = 10$ MHz separates adjacent
channels.  Within some guard bands there appears a comb of 9 leakage tones
uniformly spaced over a bandwidth of approximately 350 kHz. For example, such
combs were observed between channels 5 and 6 on Starlink 2024, between channels
5 and 6 on Starlink 1184, and between channels 7 and 8 on Starlink 2423, whereas
for other satellites no combs were observed between the same channels.
Interestingly, the between-channel combs of tones, when present, persist between
frames, whereas the mid-channel gutter leakage tones, when present, only appear
during the interval of a broadcast frame.

We suspect that the between-channel tones may be the tones tracked in
\cite{neinavaie2021exploiting,neinavaie2021acquisition} and
\cite{khalife2022first} to perform Doppler-based positioning with Starlink.  We
note that neither the mid-channel gutter tones nor the between-channel tones
appear deliberate: their presence and amplitudes are not consistent from
satellite to satellite, and the between-channel tones appear to vary in
amplitude with beam adjustments.

\subsection{Frame Layout}
As shown in Fig. \ref{fig:frame-diagram}, each frame consists of 302 intervals
of length $\Tsym = 4.4$ $\mu$s plus a frame guard interval $\Tfg$, for a total
frame period of $\Tf = 1/750$ s.  Each frame begins with the PSS, which is
natively represented in the time domain, followed by the SSS, which is formatted
as a standard 4QAM OFDM symbol.  Each frame ends with the CM1SS followed by the
CSS and the frame guard interval.  A subsequent frame may be immediately present
or not, depending on user demand.  

The known information symbols of the SSS and CSS allow a receiver to
perform channel estimation across all subcarriers at the beginning and end of
each frame, permitting within-frame interpolation.  The purpose of the CM1SS,
which arrives just before the CSS and is only partially populated with
information symbols that repeat from frame to frame, is unclear, but its
predictable elements are no doubt also useful for channel estimation.

In each frame, the OFDM symbols with index $i \in \{2,3,4,5\}$ appear to contain
header (control plane) information---likely including satellite, channel, and
modulation schedules.  We infer this from an abrupt 90-degree shift in
constellation orientation between symbol $i=5$ and $i=6$, which we interpret as
denoting a transition from header to payload symbols.  Such a shift in
orientation may be seen between the left and right panels of
Fig. \ref{fig:constellationPlot}.  The first seven or so payload symbols (from
$i=6$ to approximately $i = 12$) are sometimes 16QAM modulated, with the
remainder of the symbols 4QAM modulated.  We presume that the 16QAM symbols are
destined for users whose received SNR is sufficient to support decoding them
(about 15 dB, depending on channel coding).

The previously-mentioned $4F$-wide mid-channel gutter is present in all OFDM
symbols contained in a frame, but not in the PSS.  

\subsection{Synchronization Sequences}
The synchronization sequences are of special import for efforts to dual-purpose
Starlink signals for PNT.  As with the spreading codes of civil GNSS signals,
the synchronization sequences can be predicted by a passive (receive only) radio
and thus used to construct a local signal replica whose correlation with the
received signal yields standard pseudorange and Doppler observables, the raw
ingredients for a PNT solution.

\begin{figure}[t]
\centering
\includegraphics[width=9cm]
{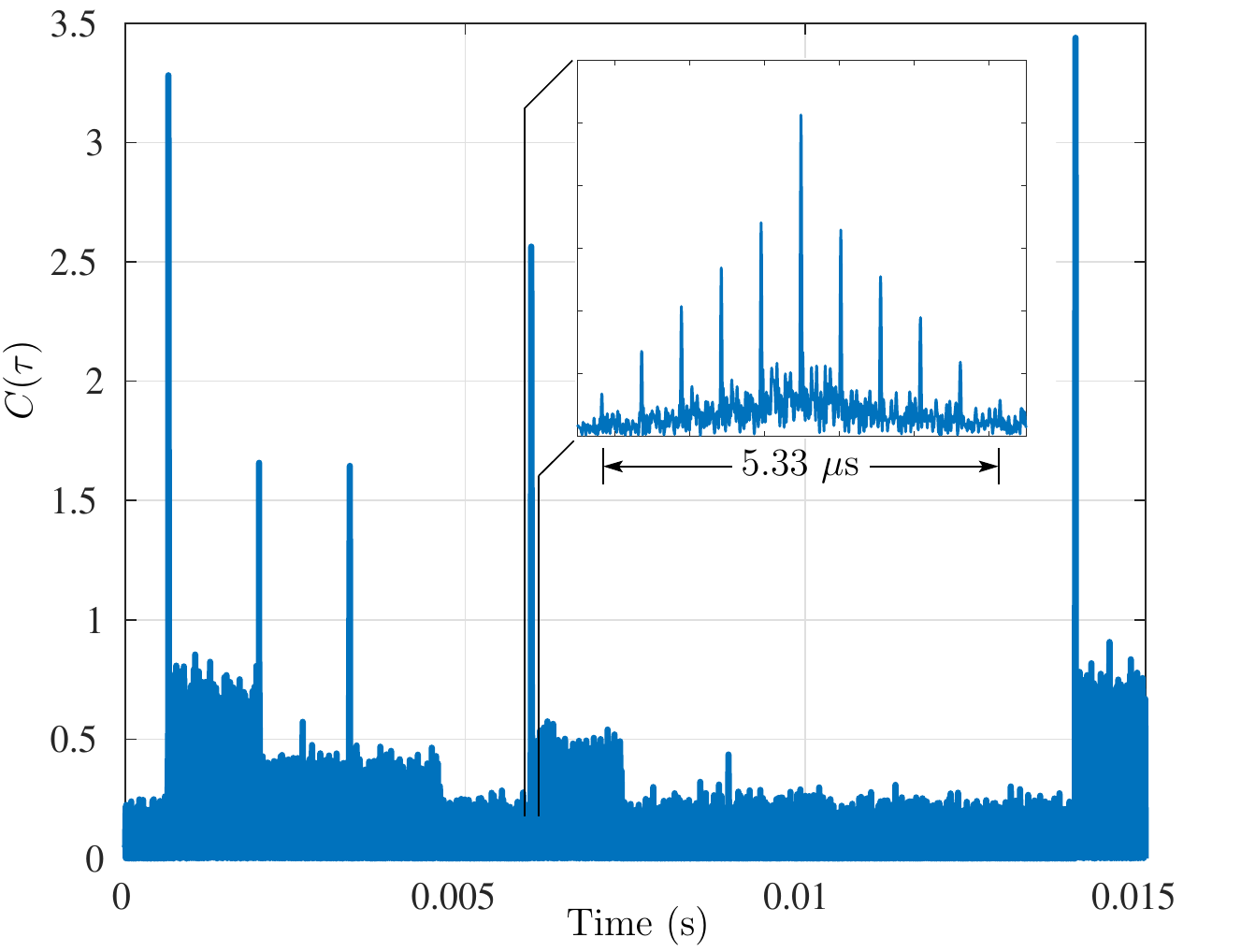}
\caption{Correlation of narrowband-mode-received Starlink data against a local
  PSS replica after Doppler compensation.}
\label{fig:frameCorrelationP}
\end{figure}

Fig. \ref{fig:frameCorrelationP} shows correlation against the PSS yielding
sharp peaks at the beginning of each frame.  The distinctive shape of the
11-tined comb shown in the figure's inset results from the repetition and
inversion of the subsequence $(p_k)_{k = N/8}^{2N/8-1}$ of which the PSS is
composed. Note that adjacent frames may have different power levels despite
being received from the same satellite and beam, evidence that the system
employs user-subset-specific power adaptation within a service cell.  Note too
the absence of frames during some intervals, which suggests that user data
demand was well below system capacity during the interval shown.  It should be
pointed out, however, that frame occupancy never dropped below 1 in 30 (one
frame every 40 ms) throughout the scores of data intervals we studied.  We
intuit that a steady stream of frames, albeit sparse, is required to support
initial network entry.  Thus, even during periods of little or no user demand,
frame arrival from each satellite will be regular and dense enough to support
opportunistic PNT.

Importantly, phase coherence is maintained throughout each frame, and the phase
relationship between the synchronization sequences appears to be constant across
frames and satellites.  This implies that time-domain representations of the
synchronization sequences (with their respective cyclic prefixes) can be
combined to extend the coherent integration interval over each frame, increasing
receiver sensitivity and observable measurement accuracy.  This technique
enables production of pseudorange and Doppler observables below -6 dB SNR, well
below the SNR required to support communication.  Thus, receivers exploiting
Starlink for PNT need not be equipped with high gain antennas and may even be
able to extract observables from satellites not servicing their cell.


Unlike GNSS spreading codes, however, the Starlink synchronization sequences are
are not unique to each satellite.  This presents a satellite assignment
ambiguity problem that must be solved combinatorially based on approximate user
location, known satellite ephemerides, and measured Doppler and frame arrival
time.

It seems clear why the PSS is composed of repeating subsequences: in high SNR
conditions, the search in Doppler and frame start time entailed by
(\ref{eq:SC_est}) for initial network entry can be made more efficient by
correlating against a single PSS subsequence and then taking the FFT of the
resulting complex accumulations with maximum modulus to refine the Doppler
estimate.  The shorter initial coherent integration interval is more forgiving
of errors in $\beta$.  If this fails due to insufficient
single-subsequence-correlation SNR, multiple subsequences can be coherently
accumulated for a slower but more sensitive search.

That the PSS is based on an m-sequence is also logical, given such sequences'
excellent autocorrelation properties \cite{dinan1998spreading}.  Encoding the
m-sequence as a series of $\pi/2$ phase shifts appears intended to reduce
spectral leakage compared to a conventional binary encoding.  The rationale for
differentially encoding the PSS is less clear.  Symmetric DPSK is known to
improve data demodulation robustness to the Doppler and timing uncertainty
common in satellite communications \cite{winters1984differential}.  But this
does not apply to coherent correlation against a known PSS (or portion thereof)
for frequency and time synchronization.  Most likely, the differential encoding
is meant to offer an additional means for trading off search sensitivity for
increased efficiency.

We were unable to identify the SSS as a canonical sequence.  Its
frequency-domain complex coefficients manifest good autocorrelation properties,
but not the constant-amplitude zero autocorrelation of m-sequences or of the
Zadoff-Chu sequences used for the PSS in LTE.  We suspect the SSS may be a
mixture of two scrambled m-sequences, as with the SSS from LTE.

\subsection{Gap to Capacity}
It is interesting to examine the Starlink signal structure in terms of its
design margins.  What balance did its designers strike in trading off data
throughput for communications reliability or cost?

\subsubsection{Spectral Occupancy} The 10-MHz guard band between channels
reduces Starlink's spectral occupancy to $\Fs/\DeltaFcenter = \frac{24}{25}$.
Leaving such a wide unused bandwidth between channels, which amounts to over 42
subcarrier intervals, suggests that Starlink intends to activate more than one
channel at a time in a given service cell and wishes to keep the costs of UTs
low by reducing their sampling rate and RF filtering requirements.

\subsubsection{OFDM Symbol Occupancy} The ratio of the useful symbol interval to
to the full OFDM symbol interval is $\T/\Tsym = \N/(\N + \Ng) = \frac{32}{33}$,
which reflects a fairly efficient design.  Compared to LTE, for which $N/\Ng$
ranges from 12.8 (more efficient) to 4 (more margin for delay spread),
Starlink's ratio is $32$.  Clearly, Starlink designers are taking advantage of
the low delay spread in the space-to-Earth channel.  Even still,
$\Tg = \Ng/\Fs = 130$ ns exceeds the worst-case 95\% root-mean-square delay
spread for the Ku-band, found in \cite{cid2015wideband} to be $\Td = 108$ ns.

\subsubsection{Frame Occupancy} One can view the frame occupancy as
$\Ndataframe\Tsym/\Tf = \frac{298}{303.03}$.  If one additionally discounts OFDM
symbols with index $i \in \{2,3,4,5\}$, which appear to contain header
information, then occupancy becomes $\frac{294}{303.03}$.  The number of OFDM
symbol intervals devoted to synchronization sequences---four every 1.33 ms---is
unusually high compared to terrestrial OFDM waveforms.  For example, LTE
transmits two synchronization sequences once every 5 ms.  By bookending each
frame with two synchronization sequences, Starlink designers ensure that UTs can
perform channel equalization and Doppler (CFO) estimation with unusually high
accuracy.  This reduces frame occupancy, but bodes well for dual-use of Starlink
signals for PNT: the greater fraction of predictable elements in each frame, the
longer a PNT-oriented receiver can coherently integrate and thus produce
pseudorange and Doppler observables at lower SNR.

\subsubsection{Channel Occupancy} Due to the $4F$-wide gutter, the channel
occupancy is at most $(N-4)/N = \frac{1020}{1024}$, but is likely somewhat
lower: Besides revealing the location of synchronization sequences, the
symbol-by-symbol frame-to-frame correlation analysis described in Section
\ref{sec:estim-synchr-sequ} suggests the presence of pilot subcarriers that are
intermittently modulated with predictable information symbols.

Another measure of channel occupancy is the subcarrier spacing $\F$.  Recall
that the number of subcarriers $N$ in $\Fs$ must be a power of two for efficient
OFDM processing, and that, \emph{ceteris paribus}, $d_\text{OFDM}$ in
(\ref{eq:ofdm_throughput}) rises with increasing $N$.  Could Starlink designers
have chosen $N = 2048$ rather than $N = 1024$, thus narrowing $\F$ by a factor
of two and increasing $d_\text{OFDM}$ by 1.54\%?  Likely so: assuming
$\N_\text{sync} = 2^{10}$ (fewer than the samples in the PSS) and SNR = 5 dB (the
threshold for 4QAM decoding assuming a benign channel and strong coding), the
constraint (\ref{eq:epsilon_approx}) could be comfortably met for $N = 2048$.

\section{Conclusions}
We have developed and applied a blind signal identification technique to uncover
the frequency- and time-domain structure of the Starlink Ku-band downlink
signal.  We further identified four synchronization sequences that can be used
to passively exploit Starlink signals for pseudorange-based positioning,
navigation, and timing (PNT), and explicitly evaluated two of these.  The
results in this paper illuminate the path to use of Starlink signals as a
backup to traditional GNSS for PNT.

\section*{Acknowledgments}
Research was sponsored by the Army Research Office under Cooperative
Agreement W911NF-19-2-0333. Additional support was provided by the
U.S. Department of Transportation (USDOT) under the University Transportation
Center (UTC) Program Grant 69A3552047138 (CARMEN), and by affiliates of the
6G@UT center within the Wireless Networking and Communications Group at The
University of Texas at Austin.  The views and conclusions contained in this
document are those of the authors and should not be interpreted as representing
the official policies, either expressed or implied, of the Army Research Office
or the U.S. Government. The U.S. Government is authorized to reproduce and
distribute reprints for Government purposes notwithstanding any copyright
notation herein.

\bibliographystyle{ieeetran} 
\bibliography{pangea}
\end{document}